\RequirePackage{rotating}
\documentclass[fleqn,usenatbib]{mnras}

\usepackage{amsmath, amsfonts, amssymb}
\usepackage{appendix}
\usepackage[caption = false]{subfig}
\usepackage[T1]{fontenc}
\usepackage{ae,aecompl}
\usepackage{graphicx}
\usepackage{rotating}
\usepackage{dcolumn}
\usepackage{caption}
\usepackage{hyperref}

\providecommand{\s}[1]{\text{#1}}

\providecommand{\e}[1]{\ensuremath{\times 10^{#1}}}

\title[The Influence of galaxy density on radio AGN power]{The relation between galaxy density and radio jet power for 1.4 GHz VLA selected AGN in Stripe 82}

\author[S. Kolwa et al.]{
S. Kolwa$^{1,2}$ \thanks{e-mail: skolwa@eso.org},
M.J. Jarvis$^{2,3},$
K. McAlpine$^{2},$
I. Heywood$^{3,4}$
\\ 
$^{1}$European Southern Observatory, Karl-Schwarzschild Str. 2, D-85748 Garching bei M\"unchen, Germany \\
$^{2}$Department of Physics and Astronomy, University of the Western Cape, Bellville 7535, South Africa \\
$^{3}$Astrophysics, Department of Physics, Keble Road, Oxford OX1 3RH \\
$^{4}$Department of Physics and Electronics, Rhodes University, PO Box 94, Grahamstown 6140, South Africa
}

\date{Accepted XXX. Received YYY; in original form ZZZ}
\pubyear{2018}
\begin{document}

\label{firstpage}
\pagerange{\pageref{firstpage}--\pageref{lastpage}}
\maketitle 
 
 %-----------------------------------------------------------------------------------------------------------------------------------------------------------
 \begin{abstract}
\noindent
Using a Karl G. Jansky Very Large Array (VLA) L-band (1-2 GHz) survey covering $\sim100$\,deg$^2$ of the Stripe 82 field, we have obtained a catalogue of 2716 radio AGN. For these AGN, we investigate the impact of galaxy density on 1.4 GHz radio luminosity ($L_{1.4}$). We  determine their close environment densities using the surface density parameter, $\Sigma_N,$ for $N=2$ and $N=5,$ which we bin by redshift to obtain a pseudo-3D galaxy density measure. Matching the radio AGN to sources without radio detections in terms of redshift, $K-$band magnitude and ($g-K$)-colour index, we obtain samples of control galaxies and determine whether radio AGN environments differ from this general population. Our results indicate that the environmental density of radio AGN and their radio luminosity are not correlated up to $z$ $\sim0.8$, over the luminosity range $10^{23} < (L_{1.4} / $W~Hz$^{-1}) < 10^{26}$. We also find that, when using a control sample matched in terms of redshift, $K-$band magnitude and colour, environments of radio AGN are similar to those of the control sample but with an excess of overdense regions in which radio AGN are more prevalent. Our results suggest that the $<1$ Mpc-scale galaxy environment plays some role in determining whether a galaxy produces a radio AGN. The jet power, however, does not correlate with environment. From this, we infer that secular processes e.g. accretion flows of cold gas to the central black-hole are more critical in fuelling radio AGN activity than radio jet power.

 \end{abstract}
 
 
%-----------------------------------------------------------------------------------------------------------------------------------------------------------
\section{Introduction}

Through recent developments in the study of galaxy evolution, it has become clear that active galactic nuclei (AGN) activity may play a critical role in the evolution of galaxies (see \citealt{heckman2014} and references therein). Semi-analytic models and hydrodynamic simulations have both introduced a requirement for physical mechanisms that terminate and/or regulate star formation in the most massive galaxies in order to reproduce the galaxy luminosity function \citep[e.g.][]{im2002,wolf2003,mauch2007,siana2008,yuan2016}. 

Many of these models have also proved that galaxies also require physical processes that would explain the relation between the black-hole mass and stellar bulge mass or host galaxy stellar mass \citep[M$_{\s{BH}}$-M$_{\rm bulge}$ or M$_{\s{BH}}$-M$_*$ relation; e.g.][]{miyoshi1995,kormendy2013} or stellar velocity dispersion \citep[M$_{\s{BH}}$-$\sigma$ relation; e.g.][]{ferrarese2000,gebhardt2000}. A popular proposition to explain both star-formation (SF) regulation and the M$_*$-$\sigma$ relation is feedback. Energy provided by the active supermassive black-hole (SMBH) of a galaxy can both heat and/or expel cold gas from the galaxy halo to quench SF in negative feedback \citep[e.g.][]{springel2005,dimatteo2005,croton2006,bower2006,hopkins2006,ciotti2010} or boost star-formation rate (SFR) in positive feedback \citep[e.g][]{ishibashi2012,zinn2013,silk2013}. Feedback from the central nucleus of the galaxy may also effect the larger scale proto-cluster or cluster environment \citep{RawlingsJarvis2004}. 

Nevertheless, the details of such feedback mechanisms is still open to question, and different mechanisms have been proposed depending on how efficiently the AGN accretes gas, leading to dichotomies that describe feedback based on the accretion efficiency as measured by the bolometric luminosity of the black-hole (BH) (L$_{\rm{bol}}$) relative to its Eddington luminosity (L$_{\rm{Edd}}$). More efficient or ``cold accretion" sources, with L$_{\rm{bol}}$/L$_{\rm{Edd}} \gtrsim$ 0.01, undergo radiative or quasar mode feedback where SF is regulated through outflows \citep[e.g.][]{fischer2010,greene2011,sturm2011,veilleux2013,spoon2013,liu2013a,liu2013b,cicone2014}. 

AGN that accrete gas less efficiently in ``hot accretion'' mode (L$_{\rm{bol}}$/L$_{\rm{Edd}}$ $\lesssim$ 0.01) via an advection dominated accretion flow \cite[ADAF;][]{heckman2014}, provide feedback mechanisms that are assumed to originate from their radio emission. The radio jets mechanically work to heat up and inflate halo gas in radio or kinetic mode feedback, where the accretion is from a hot gas reservoir in the halo of a galaxy \citep[e.g.][]{fabian2003,fabian2006,best2006}. 

Given that these two accretion mechanisms rely on the supply of cold gas, either directly from the cool gas in the intergalactic medium, or via pockets within the hot haloes surrounding individual galaxies, the environments of radio AGN should reflect this. Indeed, there have been a multitude of studies investigating the role of environment in relation to AGN activity, however, many of these give conflicting results due to the difficulty in comparing studies based on different depths, wavelengths, environment measures and redshift \citep[e.g.][]{best2004,cooper2005,sorrentino2006,tasse2011,sabater2013,
karouzos2014a,malavasi2015}.

There has, however, been a shift forward in thinking about the link between AGN and environment. This has been brought about by considering how the prevalence of AGN activity not only depends on environment but also how it depends on a galaxy's stellar and/or halo mass \citep[e.g.][]{WilliamsRottgering2015}. In addition to this, how AGN activity depends on the ongoing star formation rate (SFR) within a host galaxy, or simply the SFR within the central regions of a galaxy \citep[e.g.][]{sabater2015}. 

Using deep data across a range of wavebands, \cite{karouzos2014a} investigated how the prevalence of AGN activity depended on the local galactic environment, and found that the main driving force at high redshift may just be that the AGN preferentially reside in the most massive galaxies, rather than any link to galaxy mergers. At lower redshift, a range of studies using the Sloan Digital Sky Survey (SDSS) have started to converge. \citet{sabater2015} found that optical AGN activity at relatively low redshift is predominantly dependent on the overall gas supply into the central nucleus, and secular processes can drive the bulk of AGN activity, whereas the larger scale environment plays a secondary role by influencing this gas supply.

In a study similar to the one undertaken in this paper, \citet{ellison2015} used the SDSS and the NRAO VLA Sky Survey \citep[NVSS;][]{condon1998} to investigate the link between environment at the low-accretion regime where activity is traced by relatively low-power radio galaxies i.e. with $L_{1.4}$ in the range, 10$^{23-25}$~W Hz$^{-1}.$ The findings suggested that while such radio AGN generally reside in more evolved and massive galaxies, the primary route for fuelling the AGN comes via accretion from the surrounding medium or through minor mergers, in addition to an internal mechanism, which they attribute to stellar winds from evolved stars.

% %nature vs nurture
 Also, quite often, by examining AGN environments, more focus is placed on the physical processes occurring well outside the virial radius of the AGN's host galaxy. It is probable however that internal processes e.g. accretion, outflows, AGN feedback, magnetic field variation etc. are just as likely to play a role in triggering, sustaining or shutting off AGN activity \citep{maia2004,best2004,draper2012}. AGN studies that take into account both intrinsic galaxy properties as well as the environment provide a more complete picture. 

The radio emission measures their non-thermal component emission i.e. synchrotron radiation from AGN or star-formation activity \citep{antonucci1993}. In terms of environment, 1.4\,GHz selected AGN (z $<$ 2) with radio luminosities, $10^{21-27}$\,W~Hz$^{-1},$ are predominantly found in rich, dense cluster regions when compared to galaxies with weak or undetected radio emission \citep{malavasi2015}. Comparing optical to radio AGN (0.03 $<$ z $<$ 0.1), \citet{sabater2013} find that fractions of optical AGN decrease with density while the opposite occurs for radio AGN. Radio-loud (RL) AGN environments are also strongly clustered up to distance scales $\lesssim 1$ Mpc. This suggests that the gas medium at the scale of the dark matter (DM) halo determines both total radio jet power and the probability of a galaxy being a RL AGN \citep{donoso2010,magliocchetti2017,hale2018}.

In this paper we build on these studies and investigate the environmental density around radio AGN by combining a new deep radio survey over the SDSS Stripe 82 field with the existing deep optical data from the SDSS and near-infrared data from the UK Infrared Deep Sky Survey \citep[UKIDSS;][]{lawrence2007}. 

In this paper, we study the close environments of radio selected AGN. The paper is structured in the following way: in Section \ref{section-2}, we summarise the details of the multiwavelength dataset. We also explain the selection method for the radio AGN and three different control samples. In Section \ref{section-3}, we outline the method used to  measure environment. We detail the main results of the study in Section \ref{section-4} and discuss their implications and significance in the context of previous findings on AGN environments. Section \ref{section-5} provides a brief summary of our study and its main conclusions. Throughout this work, we have assumed $\Lambda$CDM cosmology bringing into use the fit parameters from WMAP Nine-year data (WMAP9): H$_{0} = 69.7$ km $\s{s}^{-1} \s{Mpc}^{-1}$, $\Omega_{\s{m}} = 0.237,$  $\Omega_{\s{k}} = 0.0462$ and $\Omega_{\Lambda} = 0.716$ \citep{hinshaw2013}.

%-------------
% section 2
%-------------
\section{Data and Sample Selection}\label{section-2}

\subsection{VLA 1.4 GHz Survey}
Our parent radio source catalogue is taken from a 1--2\,GHz VLA snapshot survey covering $\sim$ 100 deg$^{2}$ over the SDSS Stripe\,82 field. The survey was carried out with the CnB hybrid configuration and reaches an effective rms of $S \sim 88$\,$\mu$Jy beam$^{-1}$, allowing the detection of low surface brightness features of radio sources. Full details of the survey and the source extraction etc. can be found in \citet{heywood2016}. 

The catalogue contains 8946 5$\sigma$ detections of discrete radio sources with measurements of integrated and peak flux density. In it, each source is assigned three identifiers (IDs) to each radio detection: island, Gaussian and source IDs. In the radio survey images, a $>3\sigma$ (where $\sigma$ is the local RMS noise) detection represents contiguous emission that is identified using the island ID. Gaussian IDs are assigned to each Gaussian fit made in the islands. The source identifier PyBDSF \citep{mohan2015} then determines individual sources within each island and each of these is assigned a source ID. In this work, we make use of the source ID to determine individual sources detections. In cases where multiple detections are contiguous, a single source ID is assigned.

\subsection{SDSS DR7}\label{section-sdss}
The SDSS Stripe 82 covers an area of 275 deg$^{2}$. Here we use the SDSS Stripe 82 Co-added Galaxy Photometric Redshift Catalogue \citep{reis2012}, and impose a magnitude cut in the $g-$band of $g < 24.5.$ to reduce photometric redshift uncertainties.

\subsection{UKIDSS DR10} 
The UKIRT (United Kingdom Infrared Telescope) Infrared Deep Sky Survey (UKIDSS) has widely been considered the near-IR survey counterpart to SDSS because of its extensive survey fields that overlap with SDSS fields. We use data from the Large Area Survey (LAS) of UKIDSS which coincides spatially with Stripe\,82. UKIDSS survey data is taken in the \textit{Z,Y,J,H} and \textit{K} filters, reaching a 5$\sigma$ $K-$band Vega magnitude limit of $\sim18.2$ \citep{lawrence2007}. 

\subsection{Multi-wavelength Cross-match}
Cross-matching radio sources with optical and near-infrared counterparts is a notoriously difficult task, with many groups working on delivering automated cross-matching and characterisation algorithms \citep[e.g.][]{prescott2018}. In this paper, we adopt the relatively simple approach of a nearest neighbour to the radio centroid as it is defined for radio sources in the VLA catalogue (see \citet{heywood2016} for further details). 

However, to determine the ideal cross-match radius for multiwavelength counterpart identification, we match two catalogues and compare the distributions of their real and random cross-matches \citep{prescott2016}. We accomplish this by cross-matching both catalogues and selecting all parent catalogue (VLA) sources within a $<20$\,arcsec projected distance from their multiwavelength (SDSS or UKIDSS) counterpart. We do the same for a randomly generated catalogue containing the same number of sources as the parent catalogue, within the same field. 

\begin{figure}
 \begin{center}
  \includegraphics[width=\columnwidth]{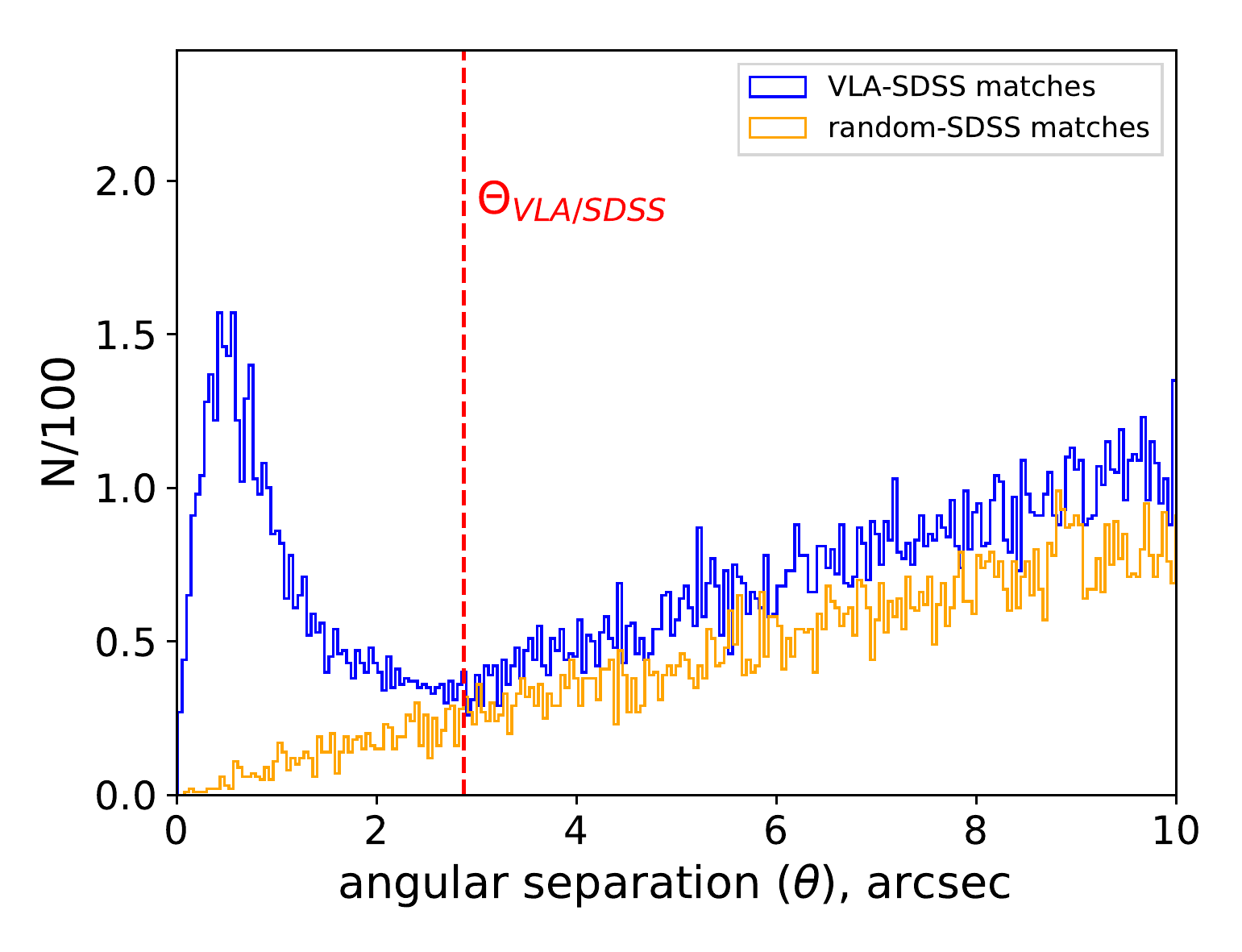}
  \caption[Cross-match radius determination for VLA/SDSS catalogue]{ Distribution of angular separations ($\theta$) between VLA sources and nearest SDSS galaxies (blue) and  between random positions and  SDSS galaxies (orange).}
  \label{fig:jvla-sdss-dist}
 \end{center}
\end{figure}

We then determine the lowest point of convergence between both distributions. The angular separation at this point indicates the boundary between projected separations that represent true or significant matches and those that are merely coincidental. The distributions are shown Fig.~ \ref{fig:jvla-sdss-dist} where we find that the two cross-matched samples converge at an angular separation of $\sim 2.9$\,arcsec. We, therefore, use this as the separation for which we are confident that the vast majority of optical identifications to the radio sources are real. We note that there will be a small fraction of false identifications, but that these should not significantly effect our results which are purely statistical.

\begin{figure}
 \begin{center}
   \includegraphics[width=\columnwidth]{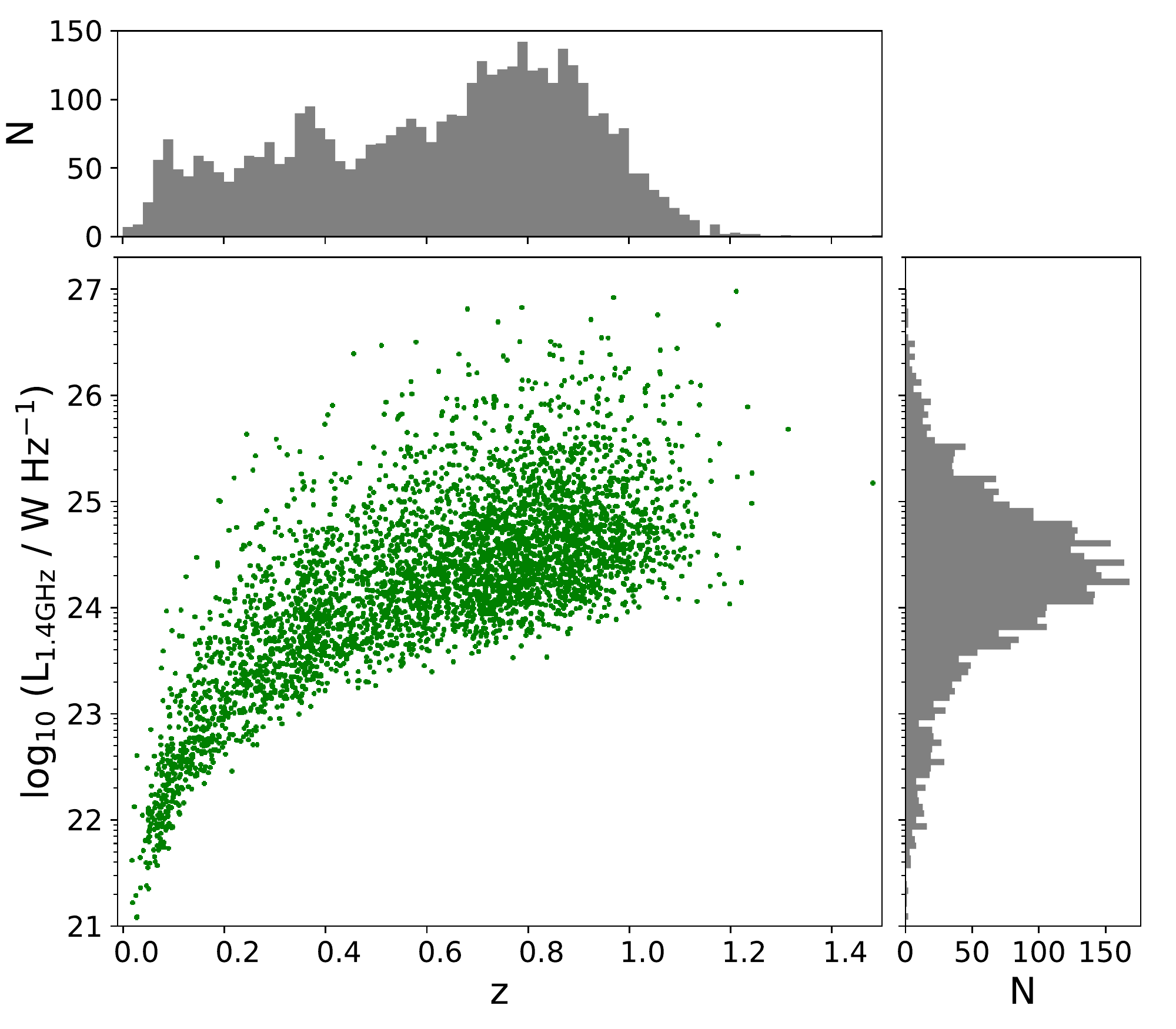}
   \caption{VLA/SDSS source 1.4 GHz radio luminosity redshift diagram. Redshift and 1.4 GHz luminosity distributions are shown as well. An increasing trend in 1.4 GHz luminosities with redshift represents a Malmquist bias in the radio data.}
   \label{fig:jvla-sdss-lum}
 \end{center}
\end{figure}

Using this radius, we find 4130 cross-identifications in SDSS, out of the total of 8946 VLA radio sources. The detection limit, in redshift, for the VLA/SDSS sample is $z\sim1.2,$ as shown by Fig.~\ref{fig:jvla-sdss-lum}. We then cross-match the SDSS position to the UKIDSS-LAS catalogue to obtain the near-infrared magnitudes for each source using a cross-match radius of 1~arcsec, yielding 2716 sources with a UKIDSS counterpart. 

\subsection{Photometric Redshifts}
We choose to only use the photometric redshifts in this study in order to ensure that the comparison of the AGN sample is comparable to that of a control sample, thus by using photometric redshifts we are able to retain the bulk of the cross-identified radio galaxies. However, photometric redshifts (z$_\s{p}$) are less precise than spectroscopic redshifts (z$_\s{s}$) and as a consequence reduce accuracy in projected distances and thus surface density measures \citep{cooper2005}. The SDSS DR7 photometric redshift catalogue \citep{reis2012} is complete to $r \sim24.5$ compared to $r \sim17.7$ for the spectroscopic redshifts \citep{york2000}. 

\begin{figure}
 \begin{center}
  \includegraphics[width=\columnwidth]{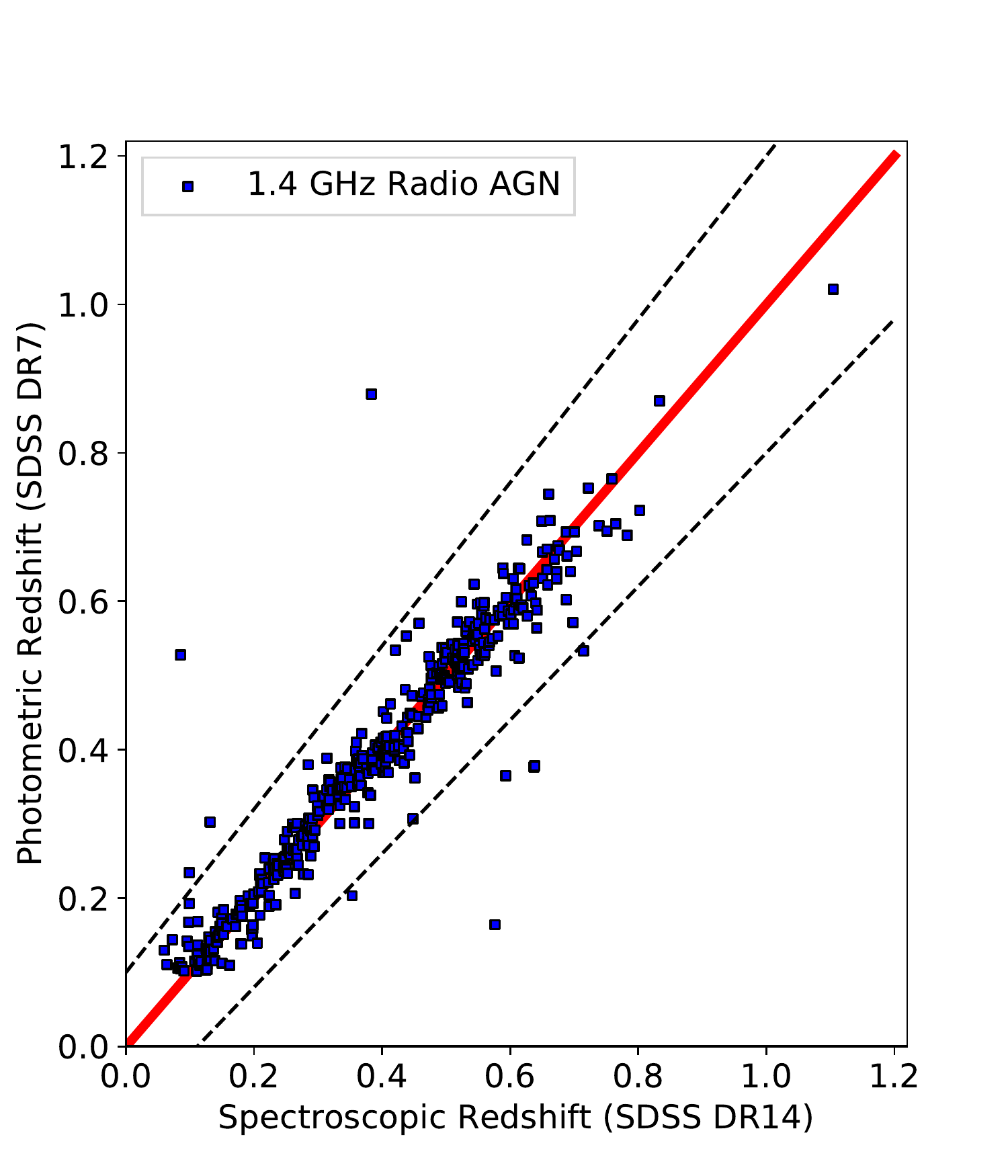}
  \caption{Comparison of SDSS spectroscopic to photometric redshifts of AGN sample. The red line represents equality between photometric and spectroscopic redshifts. The dotted lines are the upper and lower limits of the photometric redshift error.}
  \label{fig:photoz-specz}
 \end{center}
\end{figure}

We check the quality of SDSS photometric redshifts by comparing to the spectroscopic redshifts for the radio host galaxies where available. In SDSS DR14, 442 of the VLA radio sources have  spectroscopic redshifts. The photometric redshift estimates against the spectroscopic redshifts are shown in Fig.~\ref{fig:photoz-specz}. We find that 3 per cent i.e. 13 of the 442 sources are discrepant by more than $0.1(1+z_\s{s}$). Based on this, we conclude that using the photometric redshifts is not likely to bias the density measures for the VLA/SDSS sources.

% --------------
%  section 3
% --------------
\section{Analysis}\label{section-3}

\subsection{Surface Density Estimation}\label{section:env-quant}
There are several methods one can use to quantify the environmental density around galaxies. The chosen method depends on available parameters and the extent to which density is measured. In this work, we use the surface-density parameter, $\Sigma_N.$ 

\begin{equation}\label{eqn:surf-dens}
\Sigma_N = \frac{N}{\pi{d_N}^2}
\end{equation}

Here, $d_N$ is the projected distance to the $N$th nearest neighbour \citep{dressler1980,gomez2003,mateus2004}. This method requires the exclusion of foreground and background sources via either colour cuts or redshift constraints \citep{kauffman2004,cappellari2011,karouzos2014a,karouzos2014c}. The surface density is advantageous in that it yields a continuous density measure, independent of group or cluster associations \citep{cooper2005}.

To account for separation distances in the radial direction and exclude foreground and background sources, we apply a redshift constraint to Equation~\ref{eqn:surf-dens} which yields a pseudo-3D surface density. Provided that the redshift constraint or velocity interval matches the velocity dispersion of a group or cluster, the local group/cluster density is measured via $\Sigma_N.$ In general, the redshift interval is expressed as $\Delta z = \left| z_i - z_j \right| \leq \varepsilon(1 + z_i)$ for which \textit{i} and \textit{j} are the target and neighbouring galaxies, respectively. $\varepsilon$ is adjusted depending on the radial distance that the density is measured out to and importantly, depends on the accuracy of the redshifts in the sample of interest  

For our analysis we adopt a conservative value of $\varepsilon =  0.1$, i.e. setting the redshift constraint on AGN surface density $\Delta z  < 0.1(1+z_{\s{AGN}}),$ where $\Delta z = z_{\s{AGN}} - z_{\s{nn}}$ where z$_{\s{nn}}$ is the redshift of the nearest neighbour (nn). Combining this with the surface density measured in projection, we obtain a pseudo-3D AGN density measure, $\Sigma_{N,\s{AGN}}.$

We are aware that survey artefacts such as point-spread function (PSF)-like features can affect the surface density measure. As explained more fully in Section 3.2 of \citet{heywood2016}, adaptive thresholding in the source finder as well as a manual excision of visually identified artefacts have been used to ensure that the survey contains real source detections. 

Angular masks for bright stars in SDSS can bias surface density measures for either radio AGN or control galaxies that are in close proximity ($\leq150$\,arcsec i.e. the search radius for nearest neighbours) to foreground stars in the survey field. This is not likely to be an issue in the SDSS DR7 galaxy catalogue since the number of bright stars in the cross-matched Tycho-2 and SDSS DR2 fields, which overlap with Stripe 82, i.e. $\sim6\e{5}$ stars \citep{blanton2005}, is significantly lower than the total number of SDSS DR7 galaxies ($\sim130\e{5}$) \citep{reis2012} from which we select the radio AGN and control samples. Also, given that our final results are drawn from statistical means in $\Delta z = 0.1-0.2$ sized redshift bins, a minority of bright stars will not affect the mean and median environment measures significantly enough to change the results.

In addition to a redshift constraint, we apply a survey edge correction to refrain from incorrectly measuring the density for sources lying at the edge of the survey field. Measuring surface density for a source located at a projected distance $<d_N$ from the survey edge can result in the $N$th nearest neighbour being located outside the survey field which would lead to an overestimate of $d_N$.  We, therefore, remove radio sources with angular distances $<120$ \,arcsec from the edges of the combined VLA and SDSS/UKIDSS survey fields. In total, 37 radio sources meet this criterion and are removed.

We quantify galaxy environmental density using $N=2$ and $N=5$ in order to measure the close environments of the radio sources at the kpc-scale \citep{karouzos2014a,karouzos2014c}. We subdivide the radio AGN sample into four redshift intervals in order to investigate any evolution in environmental density, whilst also limiting Malmquist bias. The lowest redshift interval spans 0.1 $<$ z $<$ 0.2 to compare our results to previous low-redshift radio AGN environment studies. The remaining three redshift intervals span 0.2 $<$ z $<$ 0.8. 

We note that photometric redshift determination in SDSS is less efficient at z $\geq 0.75$ \citep{reis2012}. This may bias density measures in our sample as a reasonable portion of the VLA-SDSS sample exist at z $\geq 0.75,$ hence we apply an upper cut to the entire sample at $z$ $\sim0.8.$

\subsection{Radio Source Sample}\label{section:agn-samp}
We determine radio luminosity using flux densities from the $5\sigma$ 1--2\,GHz detections and photometric redshifts from the SDSS DR7 catalogue. The radio sources in our final sample lie at $0.1 < z < 0.8$ and we assume a radio spectral index of $\alpha = -0.7$ \citep{condon2002,sadler2002} to $k$-correct the rest-frame luminosity. 

\subsection{Control Samples}\label{section:control-samp}

In order to assess whether the radio AGN reside in higher density environments compared to the general galaxy population a control sample is required. We construct three separate control samples by matching the radio AGN host galaxies to non-radio AGN galaxies in the SDSS/UKIDSS Galaxy Catalogue in terms of (i) redshift, (ii) redshift and $K-$band magnitudes, and (iii) redshift, $K-$band magnitude and ($g-K$)-colour.

At the relatively low redshifts under investigation here, $K-$band light
is dominated by emission from low mass red stars that form the bulk of the stellar mass, therefore the $K-$band magnitude can be considered a proxy for stellar mass \citep[e.g.][]{gavazzi1996,rocca-volmerange2004}. Similarly, we can take the ($g-K$)-colour index as a proxy for the specific star formation rate (sSFR). These three control samples allow us to identify how the environment of the radio AGN may differ from the field by controlling for key properties of the galaxies. Considering them separately also allows ease of comparison with previous works which may have only accounted for one or two of the parameters considered here.

\begin{table}
\centering
\caption{1D and 2D KS-test results for sample (iii): redshift, $K-$band magnitude and ($g-K$)-colour matched field comparing the distributions for the AGN and control sample.}
\label{table:KScontrol}
\begin{tabular}{l | c c c}
\hline
 Tested samples & redshift interval & $D$ & $p$ \\
 & & & \\
  \hline
  Redshift (1D) 						& 0.1 $<$ z $<$ 0.2 		& 0.05 & 0.99 \\
   					 					& 0.2 $<$ z $<$ 0.4 		& 0.04 & 0.92 \\
   										& 0.4 $<$ z $<$ 0.6 		& 0.06 & 0.75 \\
   										& 0.6 $<$ z $<$ 0.8 		& 0.05 & 0.96 \\
                                 
    \hline
   $K-$band magnitude (1D) 				& 0.1 $<$ z $<$ 0.2 		& 0.03 & 0.99 \\
   										& 0.2 $<$ z $<$ 0.4 		& 0.02 & 0.99 \\
   										& 0.4 $<$ z $<$ 0.6 		& 0.03 & 0.99\\
    									& 0.6 $<$ z $<$ 0.8 		& 0.03 & 1.00\\                                   
  \hline
   ($g-K$)-colour (1D) 					& 0.1 $<$ z $<$ 0.2 		& 0.03 & 0.99 \\
   										& 0.2 $<$ z $<$ 0.4 		& 0.03 & 0.99 \\
   										& 0.4 $<$ z $<$ 0.6 		& 0.03 & 0.99 \\
    									& 0.6 $<$ z $<$ 0.8 		& 0.03 & 0.99 \\
  \hline
  Redshift and 							& 0.1 $<$ z $<$ 0.2 	& 0.05 & 0.99 \\
  	$K-$band magnitude (2D) 			& 0.2 $<$ z $<$ 0.4  	& 0.04 & 0.99 \\
                                    	& 0.4 $<$ z $<$ 0.6		& 0.07 & 0.85 \\
                                    	& 0.6 $<$ z $<$ 0.8		& 0.06 & 0.98 \\

  \hline
  Redshift and 							& 0.1 $<$ z $<$ 0.2 & 0.06 & 0.98 \\
  	($g-K$)-colour (2D) 					& 0.2 $<$ z $<$ 0.4 & 0.04 & 0.96 \\
                                    	& 0.4 $<$ z $<$ 0.6 & 0.06 & 0.93 \\
                                    	& 0.6 $<$ z $<$ 0.8 & 0.04 & 0.99 \\
	
  \hline	
  ($g-K$)-colour and 						& 0.1 $<$ z $<$ 0.2 & 0.05 & 0.99 \\
  $K-$band magnitude (2D) 				& 0.2 $<$ z $<$ 0.4 & 0.03 & 0.99 \\
                                    	& 0.4 $<$ z $<$ 0.6 & 0.03 & 0.99 \\
                                    	& 0.6 $<$ z $<$ 0.8 & 0.04 & 0.99 \\
  \hline
  \end{tabular}
\end{table}

\begin{figure}
 \begin{center}
   \includegraphics[width=\columnwidth]{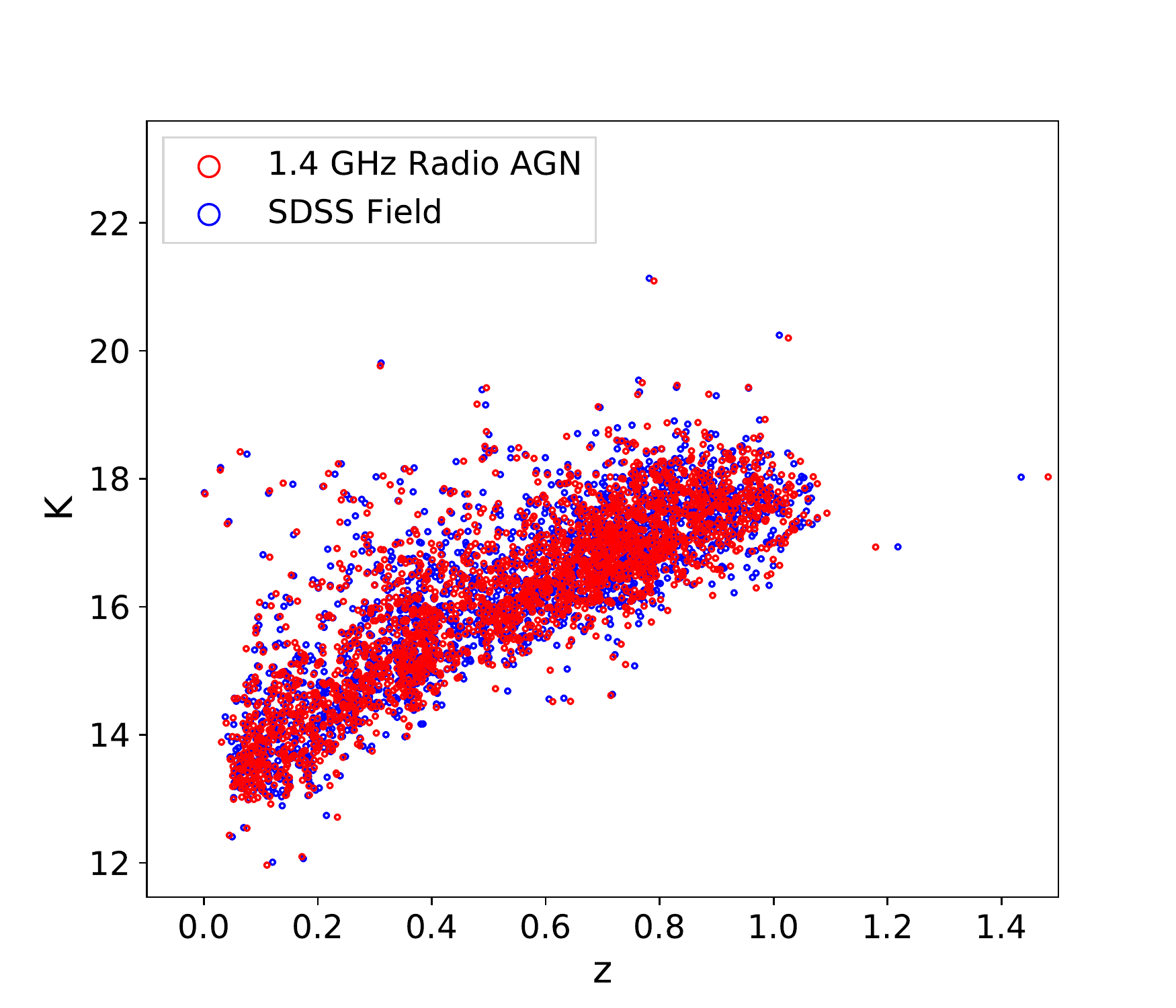}
   \caption{Petrosian $K-$band magnitude as a function of photometric redshift for radio AGN and control galaxies in sample (ii).}
   \label{fig:K-z-control}
 \end{center}
\end{figure}

\begin{figure}
  \begin{center}
    \includegraphics[width=\columnwidth]{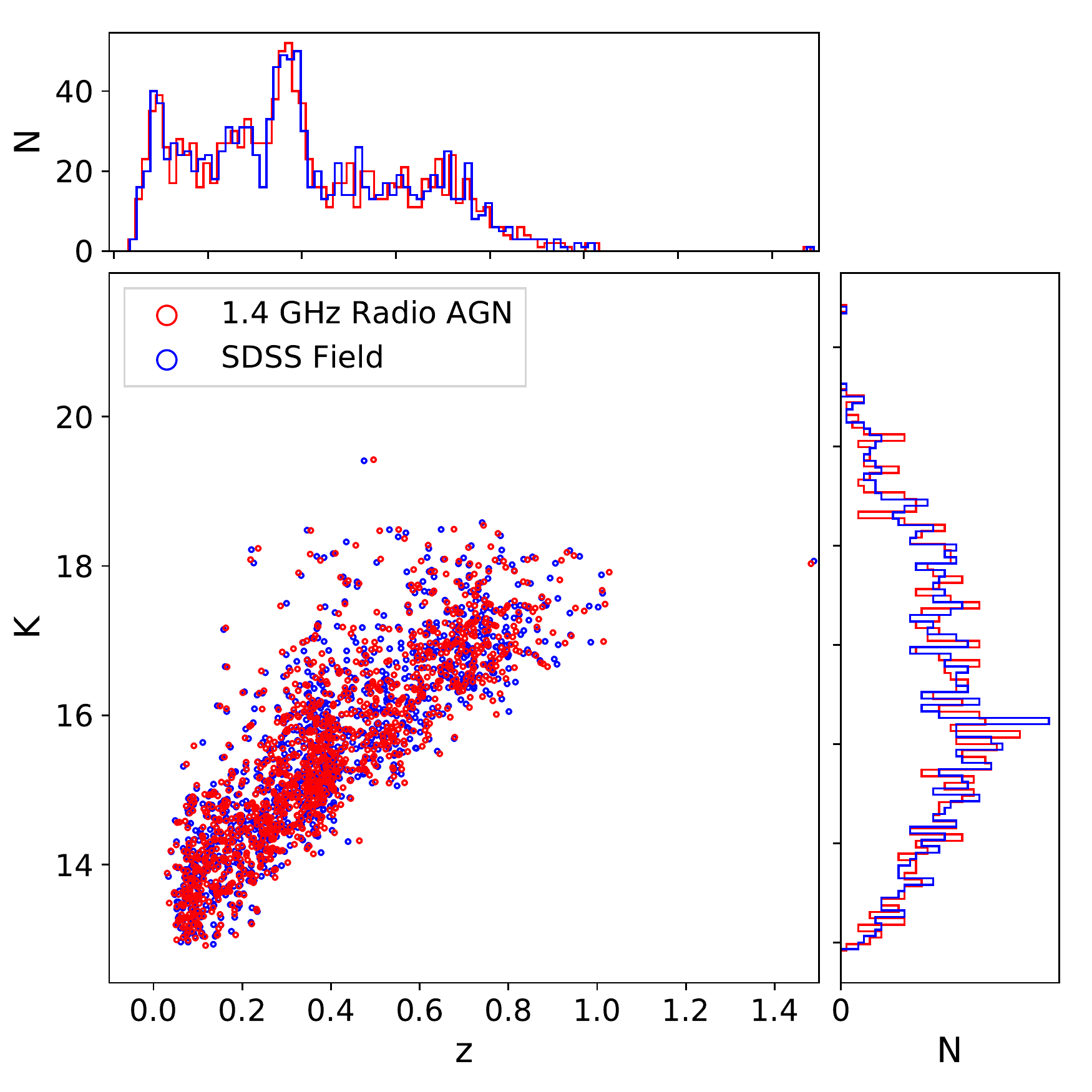}
    \caption{Petrosian $K-$band magnitude as a function of photometric redshift for radio AGN and redshift, M$_*$ and sSFR matched SDSS field galaxies in sample (iii). The \textit{1.4 GHz radio AGN} sample is a combination of VLA, SDSS and UKIDSS data. The SDSS field sources (control sample) combined SDSS and UKIDSS.}
    \label{fig:K-z-supercontrol}
  \end{center}
 \end{figure}

\begin{figure}
 \begin{center}
   \includegraphics[width=\columnwidth]{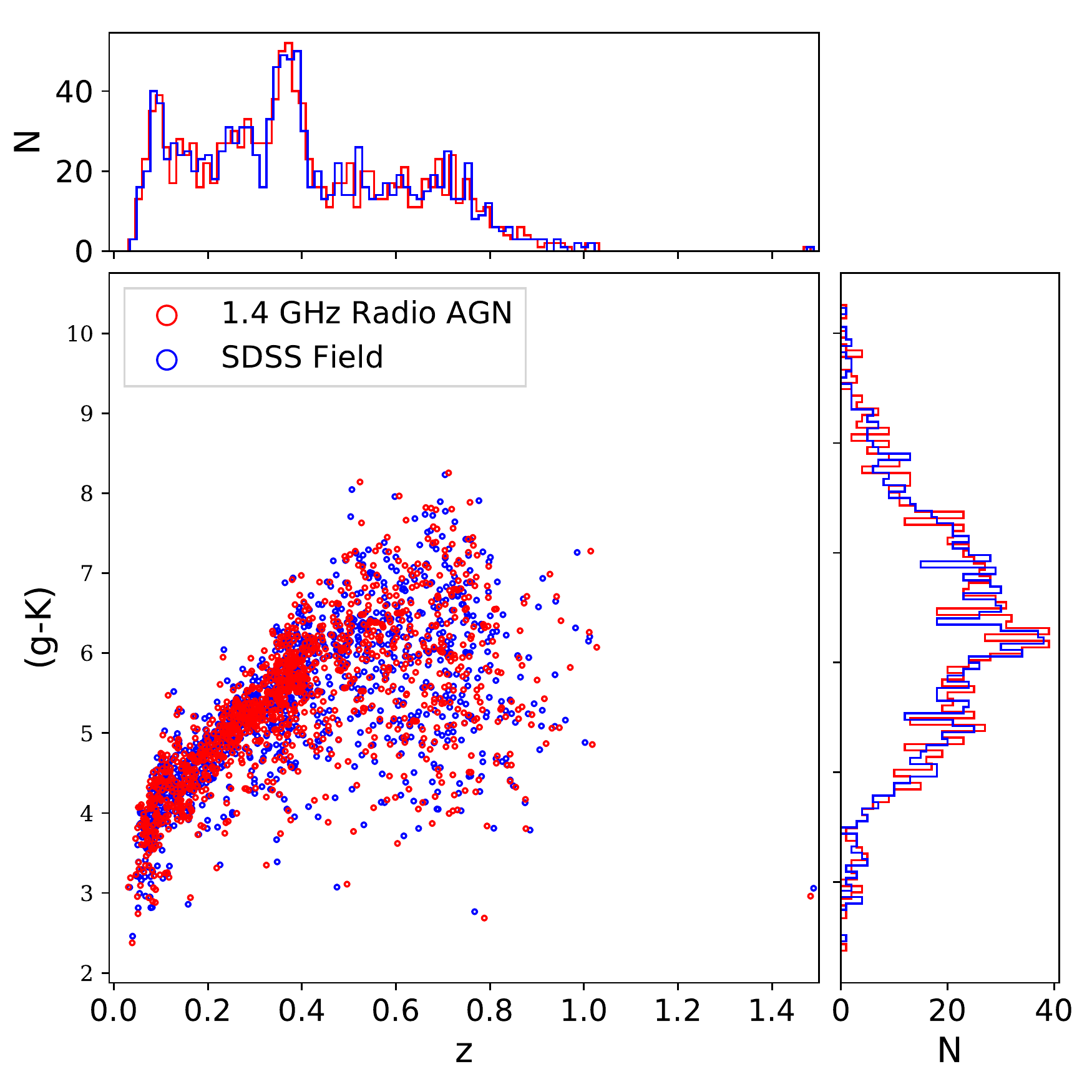}
   \caption{($g-K$)-colour as a function of photometric redshift for radio AGN and control galaxies in sample (iii).}
   \label{fig:g-K-z-supercontrol}
 \end{center}
\end{figure}

We match the radio and control sources according to the following criteria: $\Delta z \leq 0.02(1+z_{\s{AGN}}),$ $\Delta K \leq 0.05,$ and $\Delta(g-K) \leq 0.1$, where $\Delta$ denotes the difference between the AGN and control sample galaxies in the given observational quantity. The distributions are shown in Figs.~\ref{fig:K-z-control}, \ref{fig:K-z-supercontrol} and \ref{fig:g-K-z-supercontrol}. We perform 1- and 2-D Kolmogorov-Smirnov tests (KS-tests) to check that the distributions in $z$, $K$ and ($g-K$) are indistinguishable between the radio AGN hosts and the control samples. The results of this are shown in Table~\ref{table:KScontrol}. We find that the underlying distributions are indistinguishable for all our sub-samples.

Each radio source in the combined VLA/SDSS/UKIDSS catalogue is matched to a single control galaxy located at an angular separation $>60$\,arcsec from a radio AGN following the criteria in (i), (ii) and (iii). This angular distance constraint is set to reduce the likelihood of the radio source and control galaxy environments overlapping. 

As for the radio AGN, we measure the environmental density around each control galaxy in redshift slices of $| z_{\s{cont}} - z_{\s{nn}}| = 0.1(1+z_{\s{cont}})$, enabling us to compare the radio AGN environments directly with the control sample environments.

\subsection{AGN environment relative to the field} 
The pseudo-3D density of a radio AGN's environment is denoted by $\Sigma_{N,\s{AGN}}$ and the environmental density for a set of control galaxies matched to a radio AGN is denoted by $\Sigma_{N,\s{field}}.$

To determine the density of radio AGN environments relative to the field, we take a ratio of these two density measures. Hence, we define $\Sigma_{N,\s{AGN}}/\Sigma_{N,\s{field}}$ as the relative density i.e. $\Sigma_{N,\s{R}}.$ We give this density as both a mean and a median, along with the associated error on the mean, and 16th and 84th percentiles for the median uncertainties. 

Using this, we examine the relation between relative density and $L_{1.4}$ for the radio AGN sample in the 4 redshift intervals spanning 0.1 $<$ z $<$ 0.8. The mean and median relative densities as a function of $L_{1.4}$ for the radio AGN in the redshift slices tell us how jet power correlates with environment density. Spearman's rank correlation parameters $\rho$ and $p$ determine this correlation. For $p < 0.05,$ we reject the null hypothesis that no correlation exists. For environment density as a function of 1.4 GHz radio power the correlation test results are shown in Table \ref{table:correlation_parameters}. Mean and median density are significant above the $3\sigma$ level in each $L_{1.4}$ bin. Mean and median densities are referred to as $\langle \Sigma_{N} \rangle$ and $\widetilde{\Sigma_{N}}$, respectively. We use two-sample 1D and 2D KS tests to measure the similarity of AGN and control sample environment density distributions. 

%$We also use them as goodness of matching tests for redshift, $K$-band magnitude and ($g-K$) colour between AGN and control samples. Two samples that are similar or more likely to be drawn from the same parent distribution for p $>0.05.$ 

\begin{table}
\begin{center}
 \caption{Spearman's rank correlation test parameters for the relative surface density-$L_{1.4}$ relation for the redshift intervals considered i.e. control sample (i)}
 \label{table:correlation_parameters}
  \begin{tabular}{ l l | r r }
  \hline
 & redshift interval & $\rho$ & $p$ \\
\hline
  $\frac{\Sigma_{2,{\rm AGN}}}{\Sigma_{2,{\rm field}}}$ 	& 0.1 $<$ z $<$ 0.2 & 0.03 & 0.74	\\
   						& 0.2 $<$ z $<$ 0.4 & 0.00 & 0.99	\\
   						& 0.4 $<$ z $<$ 0.6 & 0.01 & 0.91	\\
   						& 0.6 $<$ z $<$ 0.8 & 0.01 & 0.86	\\         
  \hline                                                                                            
  $\frac{\Sigma_{5,{\rm AGN}}}{\Sigma_{5,{\rm field}}}$ 	& 0.1 $<$ z $<$ 0.2  & 0.16 & 0.04 \\
   						& 0.2 $<$ z $<$ 0.4  & 0.06	& 0.20 \\     
  	 					& 0.4 $<$ z $<$ 0.6  & 0.01 & 0.88  \\         
  						& 0.6 $<$ z $<$ 0.8  & 0.00 & 0.98 	\\
  \hline
  \end{tabular}
 
 \end{center}
\end{table}

\subsubsection{Redshift-matched Control Sample}

\begin{figure*}
 \begin{center}
  \includegraphics[width=0.95\columnwidth]{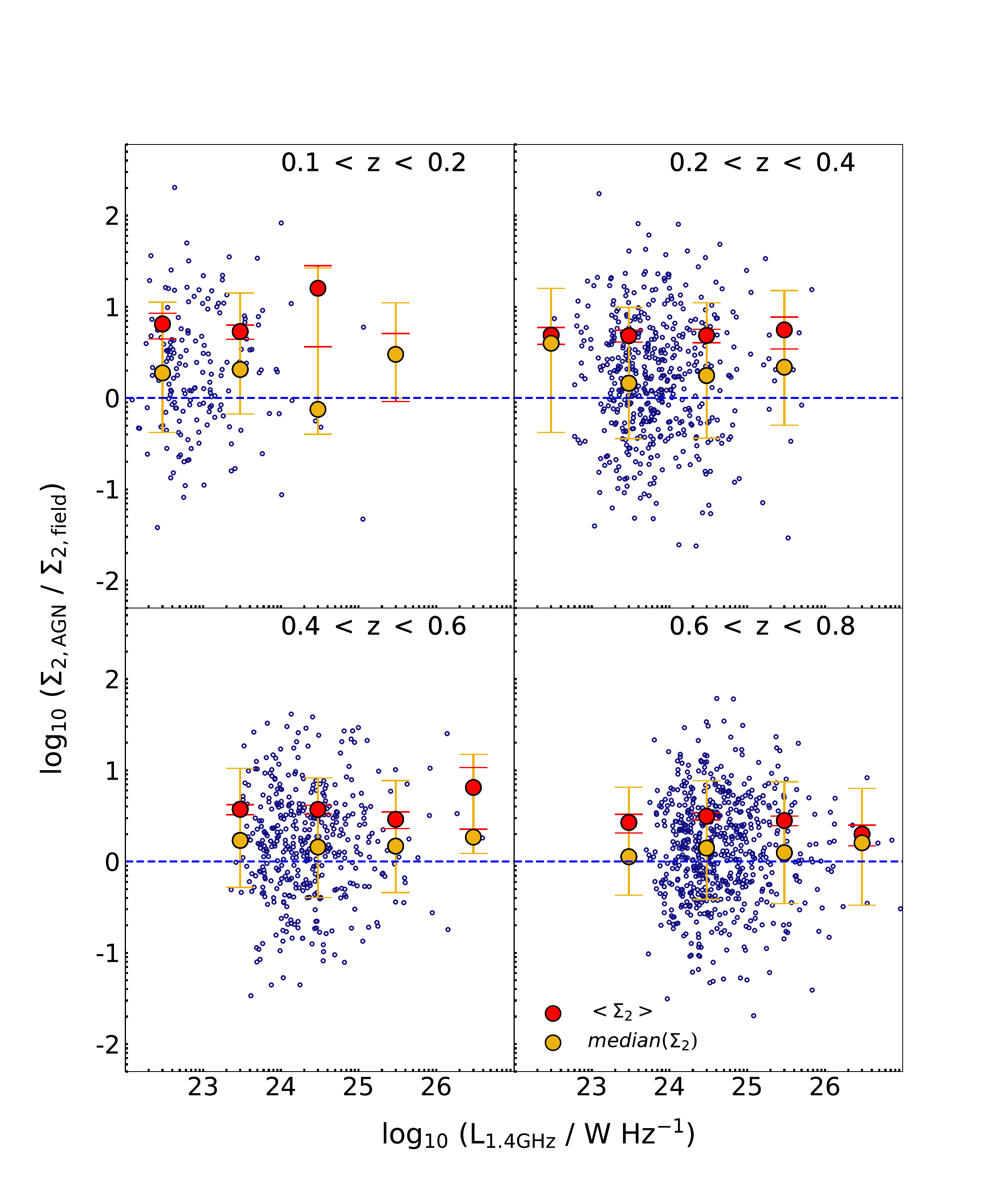}
\includegraphics[width=0.95\columnwidth]{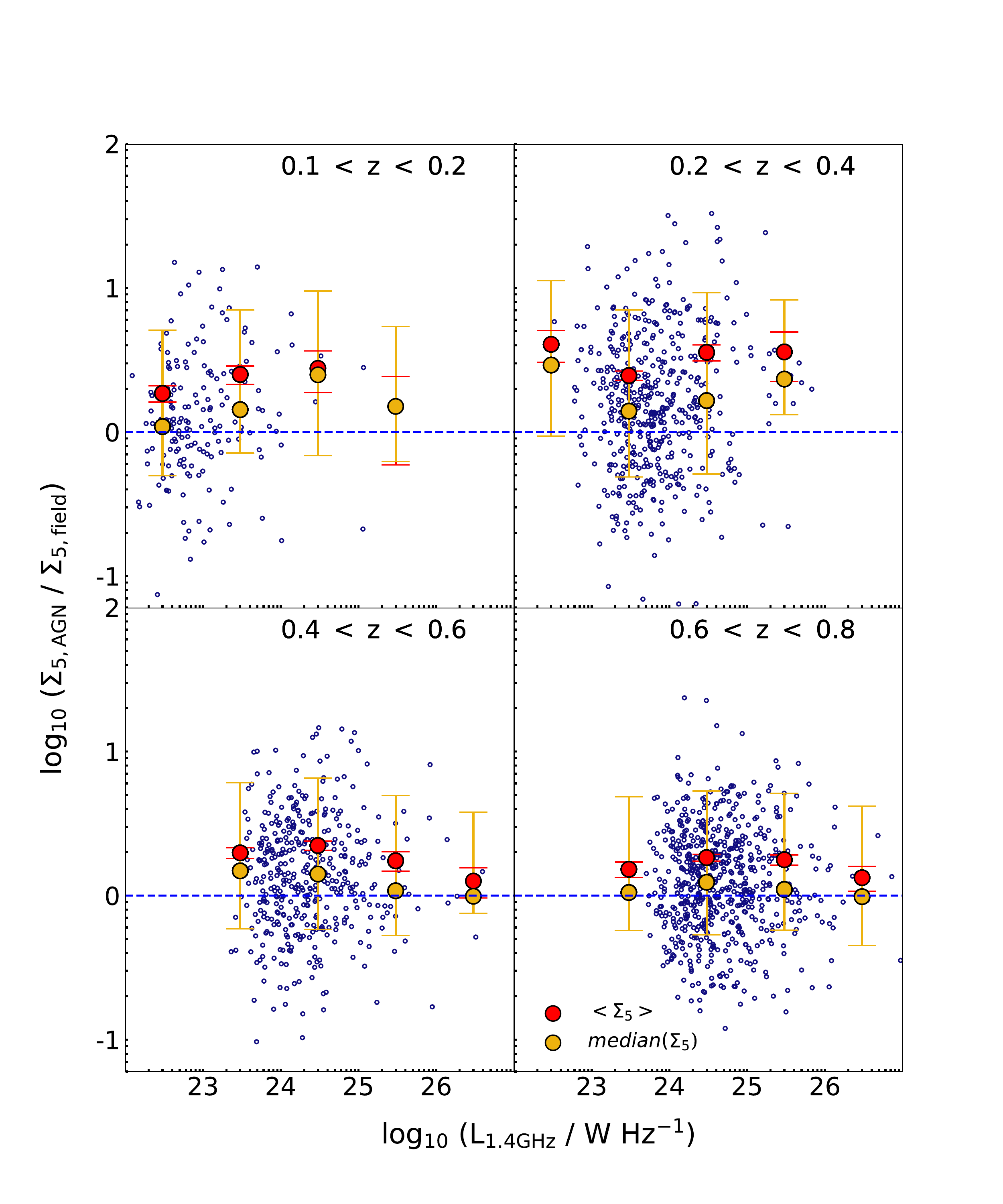}
  \caption{Mean and median $\Sigma_{2,\s{R}}$ ($left$) and $\Sigma_{5,\s{R}}$ ($right$) for the AGN relative to field densities as a function of $L_{1.4}$ for the redshift-matched control sample in 4 redshift intervals. Relative density is denoted by the ratio between AGN and field densities. Mean density ($\langle \Sigma_{2,\s{R}} \rangle,$ red) and median ($\Sigma_{2,\s{R}}$, yellow) per $L_{1.4}$ interval are shown. The dashed line denotes the line of equality for the AGN and control sample environments.}
  \label{fig:random-env_L-radio}
\end{center}
\end{figure*}

% \begin{figure*}
%  \begin{center}
%   \includegraphics[width=0.95\columnwidth]{plots/random_env_2_histogram.pdf}
%   \includegraphics[width=0.95\columnwidth]{plots/random_env_5_histogram.pdf}
%   \caption{Distributions of $\Sigma_{2,\s{R}}$ ($left$) and $\Sigma_{5,\s{R}}$ ($right$) of AGN and  redshift-matched field samples are shown in blue and green, respectively in 4 redshift intervals.}
%   \label{fig:random-env_L_radio_hist}
%  \end{center}
% \end{figure*}

\begin{table}
\centering
\caption{$L_{1.4}$-bin KS-test results for control sample (i): redshift-matched.}
\label{table:sample1b}
\begin{tabular}{ c | c c c }
\hline
  & redshifts & $D$ & $p$ \\
  & & & \\
  \hline
   $\frac{\Sigma_{2,{\rm AGN}}}{\Sigma_{2,{\rm field}}}$ 				& 0.1 $<$ z $<$ 0.2  &	0.242  &  1.4\e{-5} \\	
   						 	& 0.2 $<$ z $<$ 0.4  &	0.248  &  1.5\e{-14} \\	
   							& 0.4 $<$ z $<$ 0.6  & 	0.259  &  7.3\e{-14} \\	
   							& 0.6 $<$ z $<$ 0.8  &  0.156  &  1.0\e{-7} \\	
  \hline
  $\frac{\Sigma_{5,{\rm AGN}}}{\Sigma_{5,{\rm field}}}$ 				& 0.1 $<$ z $<$ 0.2  &	0.164  &  9.8\e{-3} \\	
   							& 0.2 $<$ z $<$ 0.4  & 	0.276  &  8.3\e{-18} \\	
   							& 0.4 $<$ z $<$ 0.6  & 	0.231  &  3.8\e{-11} \\ 	
   							& 0.6 $<$ z $<$ 0.8  &    0.140  &  2.7\e{-6} \\	
  \hline
  \end{tabular}
\end{table}

First, we compare the relative density between the AGN and the control sample matched in redshift only. We find that the mean relative density is $>1$ in every redshift interval at $>3\sigma$ significance. In terms of the absolute overdensity, we find that the environments of AGN significantly exceed those of non-AGN (Fig.~\ref{fig:random-env_L-radio}). However, we do not find a significant correlation between the relative environmental density at the radio luminosity of the AGN using a Spearman's rank correlation test $p$-values (Table~\ref{table:correlation_parameters}) pointing to an absence of correlation between $L_{1.4}$ and radio AGN environment density. 

The KS-tests, presented in Table~\ref{table:sample1b}, show that the environments of the AGN are statistically different to those of redshift-matched control samples. However, by only matching the AGN and control samples in terms of redshift we are neglecting the known dependence of radio luminosity on host galaxy mass \citep[e.g.][]{McLure2004,WilliamsRottgering2015}, and possibly colour. We, therefore, investigate this in the following sections. 

\subsubsection{Redshift and $K-$band magnitude-matched Control Sample}

\begin{figure*}
 \begin{center}
  \includegraphics[width=0.95\columnwidth]{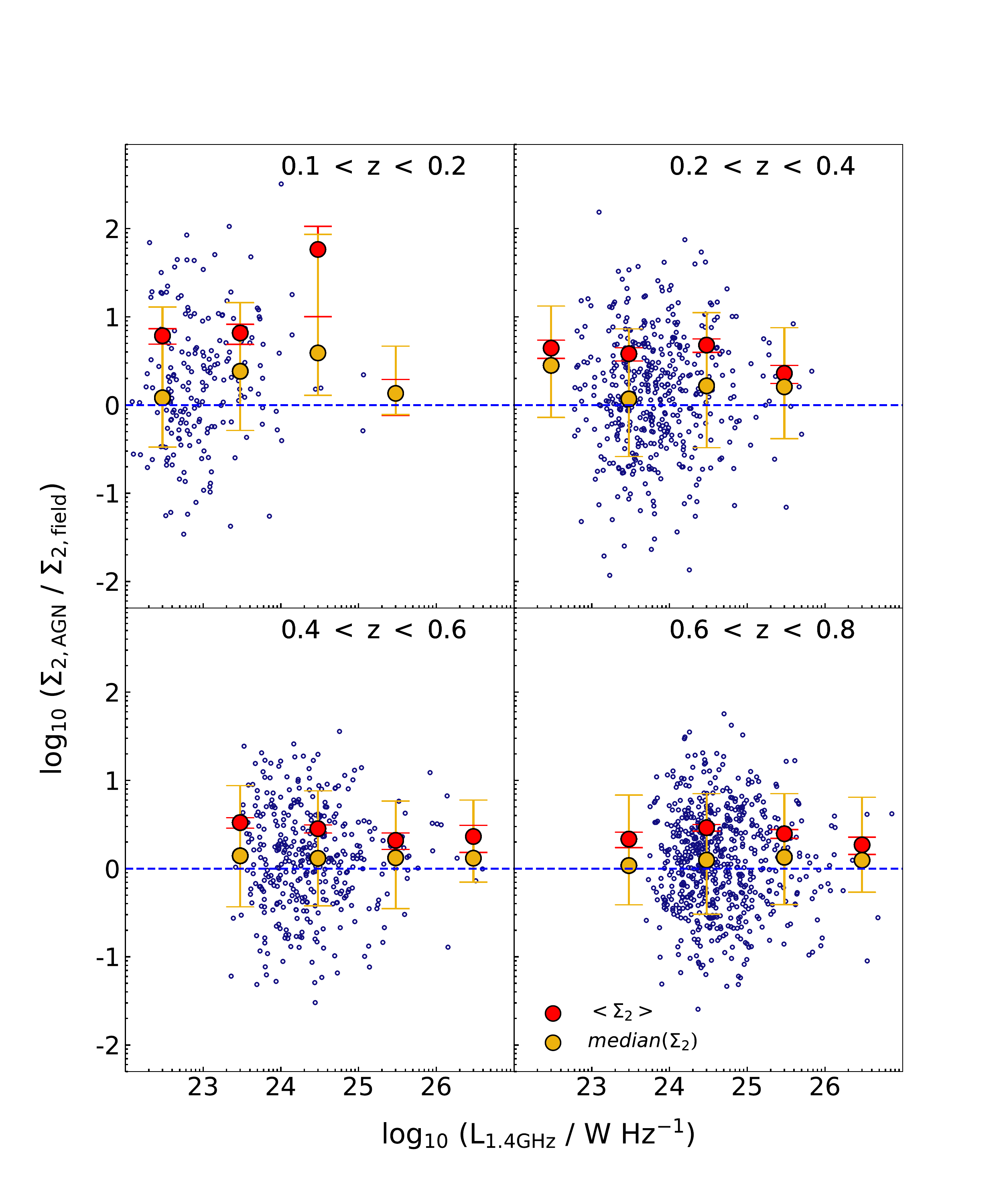}
  \includegraphics[width=0.95\columnwidth]{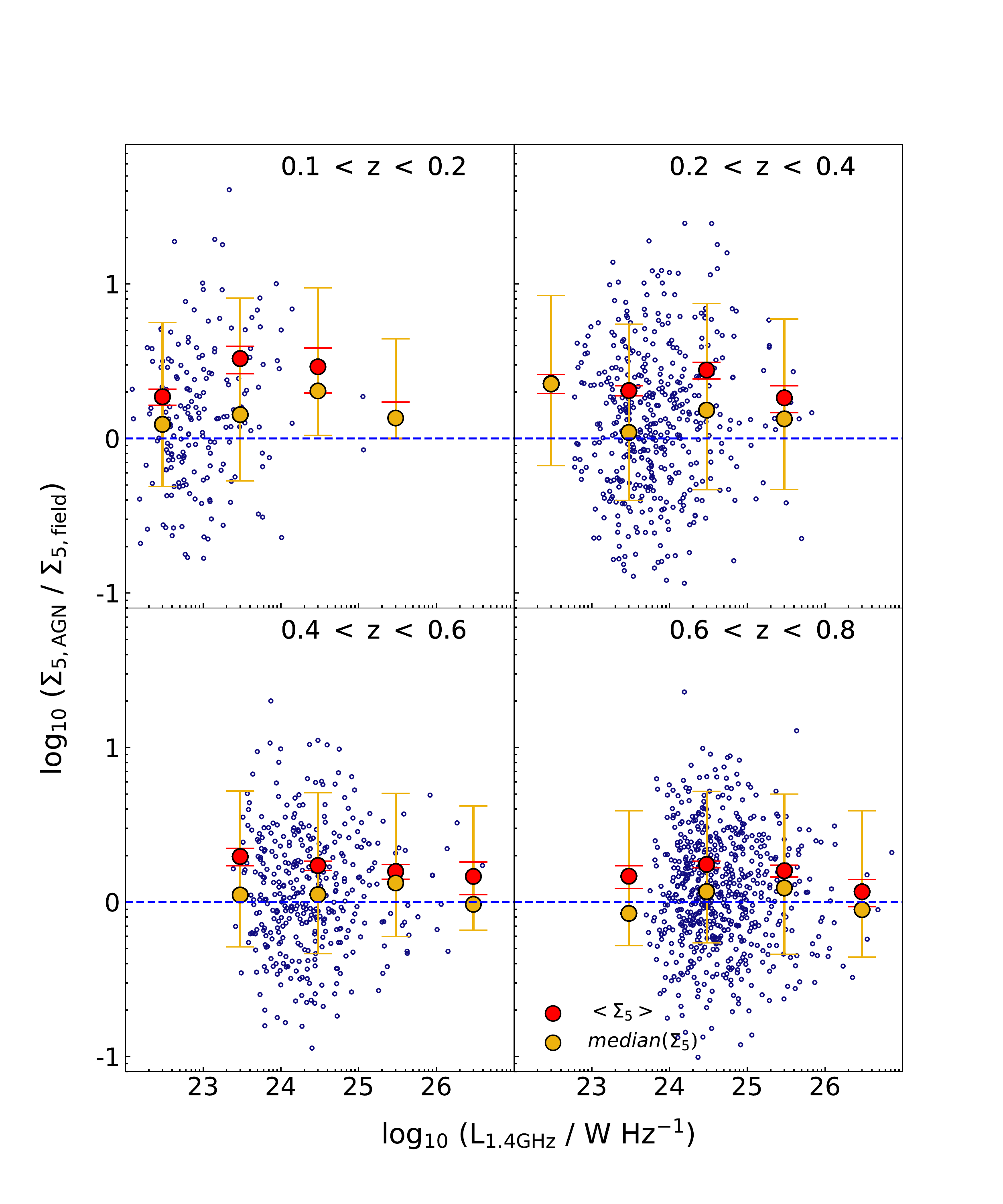}
  \caption{Same as Fig.~\ref{fig:random-env_L-radio} but for the redshift and $K-$band magnitude matched AGN and field galaxies as a function of $L_{1.4}$.}
  \label{fig:control-env_L_radio}
 \end{center}
\end{figure*}

% \begin{figure*}
%  \begin{center}
%   \includegraphics[width=0.95\columnwidth]{plots/control_env_2_histogram.pdf}
%   \includegraphics[width=0.95\columnwidth]{plots/control_env_5_histogram.pdf}
%   \caption{Same as Fig.~\ref{fig:random-env_L_radio_hist} except for the redshift and $K-$band magnitude matched AGN and field galaxies.}
%   \label{fig:control-env_L_radio_hist}
%  \end{center}
% \end{figure*}

In sample (ii), the AGN and control sources are matched in redshift and $K-$band magnitude, which is related to the stellar mass of the galaxy. As depicted in Fig.~\ref{fig:K-z-supercontrol}, the $z$ and $K-$band magnitude source matching is successful given the overlapping distributions in AGN and control distributions.
% Correlation test parameters for the density-$L_{1.4}$ relation of the redshift and $K-$band magnitude matched sample are shown in Table~\ref{table:correlation_parameters}.
We find that the mean relative density is consistently $>1$ in both $\Sigma_{2,\s{R}}$ and $\Sigma_{5,\s{R}}$ for all redshift intervals and $L_{1.4}$ intervals, as shown by Fig. \ref{fig:control-env_L_radio}. 

%Densities of radio AGN exceed those of non-AGN by factors up to $\sim60$ in $\Sigma_2$ and up to $\sim6$ in $\Sigma_5.$ 

According to the KS-tests (Table~\ref{table:sample2b}), the control sample is drawn from a different underlying distribution to that of the AGN for both $\Sigma_{2,\s{R}}$ and $\Sigma_{5,\s{R}}$, in all redshift bins. However, we find that the significance of this is reduced compared to the redshift-only control sample, suggesting that the $K-$band magnitude, or more physically, the stellar mass plays a crucial role in the measurement of the environmental density. This is expected as a multitude of studies have shown that more massive galaxies are more highly clustered, and thus reside in higher mass dark matter haloes \citep[e.g.][]{Norberg2002,Zehavi2011}. Given that higher mass haloes tend to all contain a high number of satellites \citep{mccracken2015,Hatfield2016}, this finding is in agreement with the accepted view offered by hierarchical galaxy formation models.

\begin{table}
\centering
\caption{$L_{1.4}$-bin KS-test results for control sample (ii): redshift and $K-$band magnitude-matched field.}
\label{table:sample2b}
\begin{tabular}{c | c c c}
\hline
  & redshifts & $D$ & $p$ \\
  & & & \\
  \hline
   	$\frac{\Sigma_{2,{\rm AGN}}}{\Sigma_{2,{\rm field}}}$	& 0.1 $<$ z $<$ 0.2 	&	0.256  &  1.4\e{-6} \\
   							& 0.2 $<$ z $<$ 0.4 	& 	0.164  &  2.3\e{-6} \\
   							& 0.4 $<$ z $<$ 0.6 	&	0.157  &  3.7\e{-5} \\
   							& 0.6 $<$ z $<$ 0.8 	&	0.116  &  2.2\e{-4} \\
  \hline
   $\frac{\Sigma_{5,{\rm AGN}}}{\Sigma_{5,{\rm field}}}$	& 0.1 $<$ z $<$ 0.2 	& 	0.175  &  3.1\e{-3} \\
   							& 0.2 $<$ z $<$ 0.4 	& 	0.144  &  5.4\e{-5} \\
   							& 0.4 $<$ z $<$ 0.6 	& 	0.124  &  2.4\e{-3} \\
   							& 0.6 $<$ z $<$ 0.8 	& 	0.114  &  3.0\e{-4} \\               
  \hline
  \end{tabular}
\end{table}

\subsubsection{Redshift, $K-$band magnitude and ($g-K$)-matched Control Sample}

\begin{figure*}
 \begin{center}
  \includegraphics[width=0.95\columnwidth]{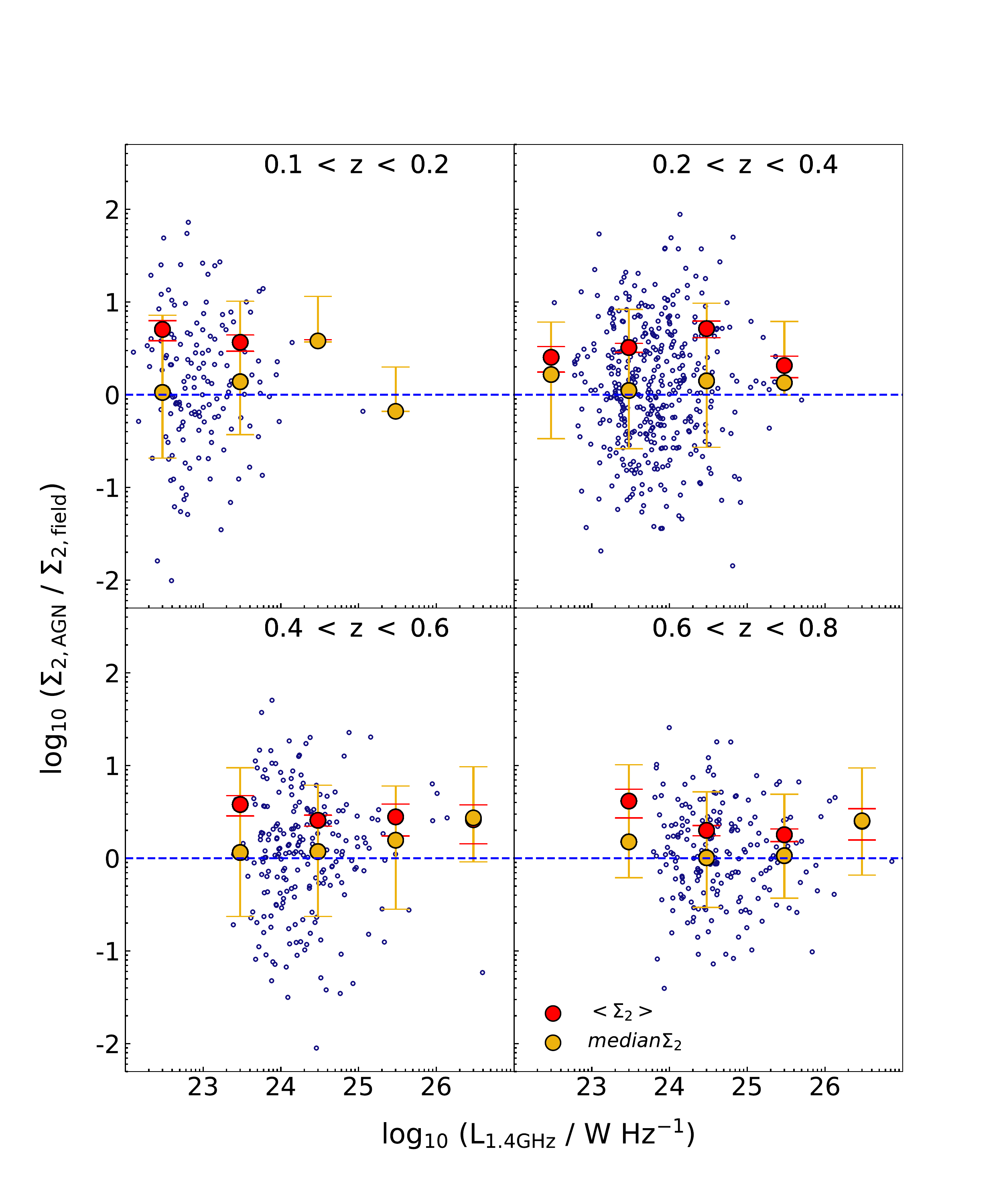}
  \includegraphics[width=0.95\columnwidth]{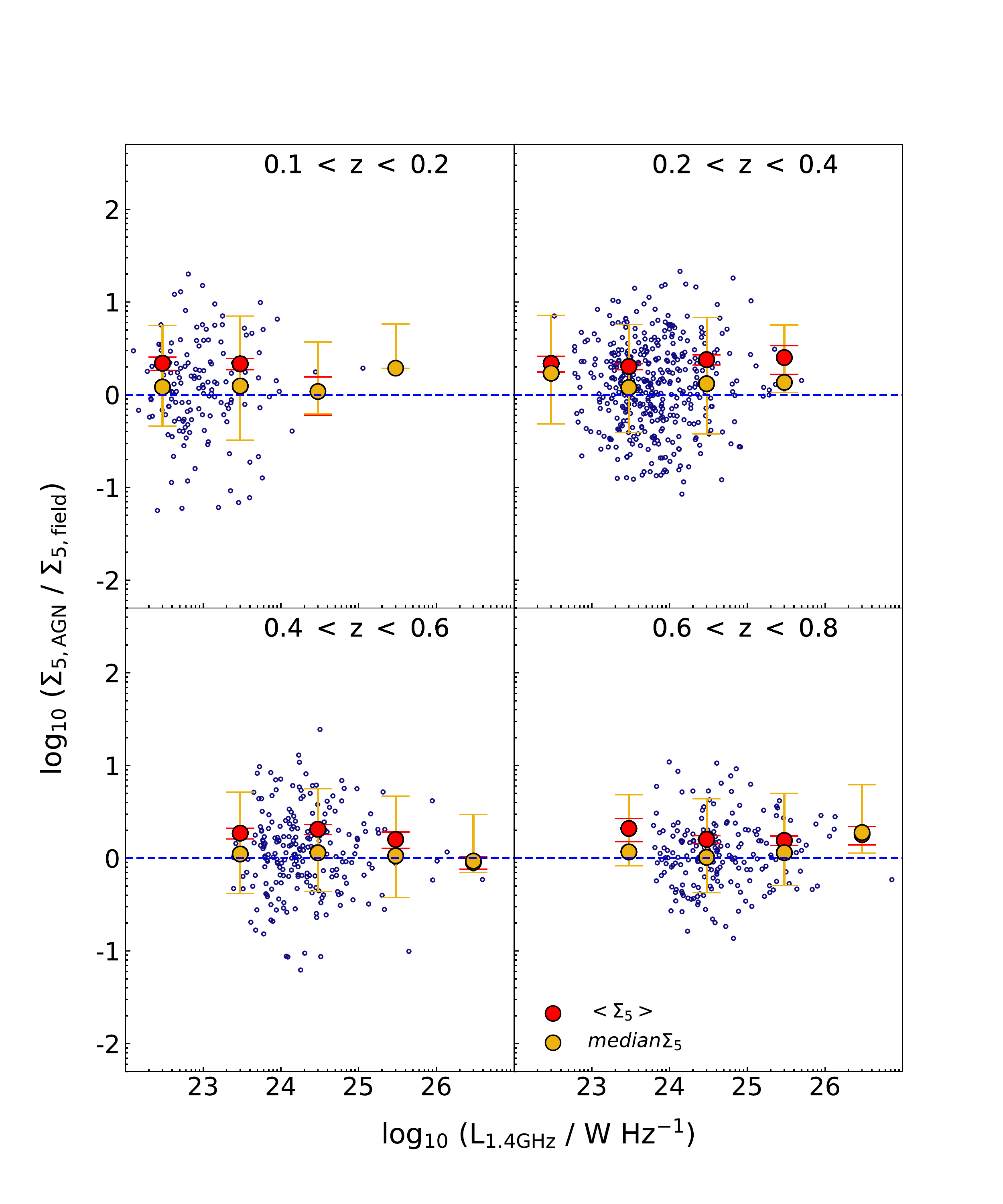}
  \caption{Same as Fig.~\ref{fig:random-env_L-radio} but for the redshift, $K-$band magnitude and ($g-K$)-colour matched AGN and field galaxies as a function of $L_{1.4}$. }
  \label{fig:supercontrol-env_L-radio}
 \end{center}
\end{figure*}

\begin{figure*}
\centering
 \subfloat[$\Sigma_2$ distributions for AGN and control samples]{\includegraphics[width=0.95\columnwidth]{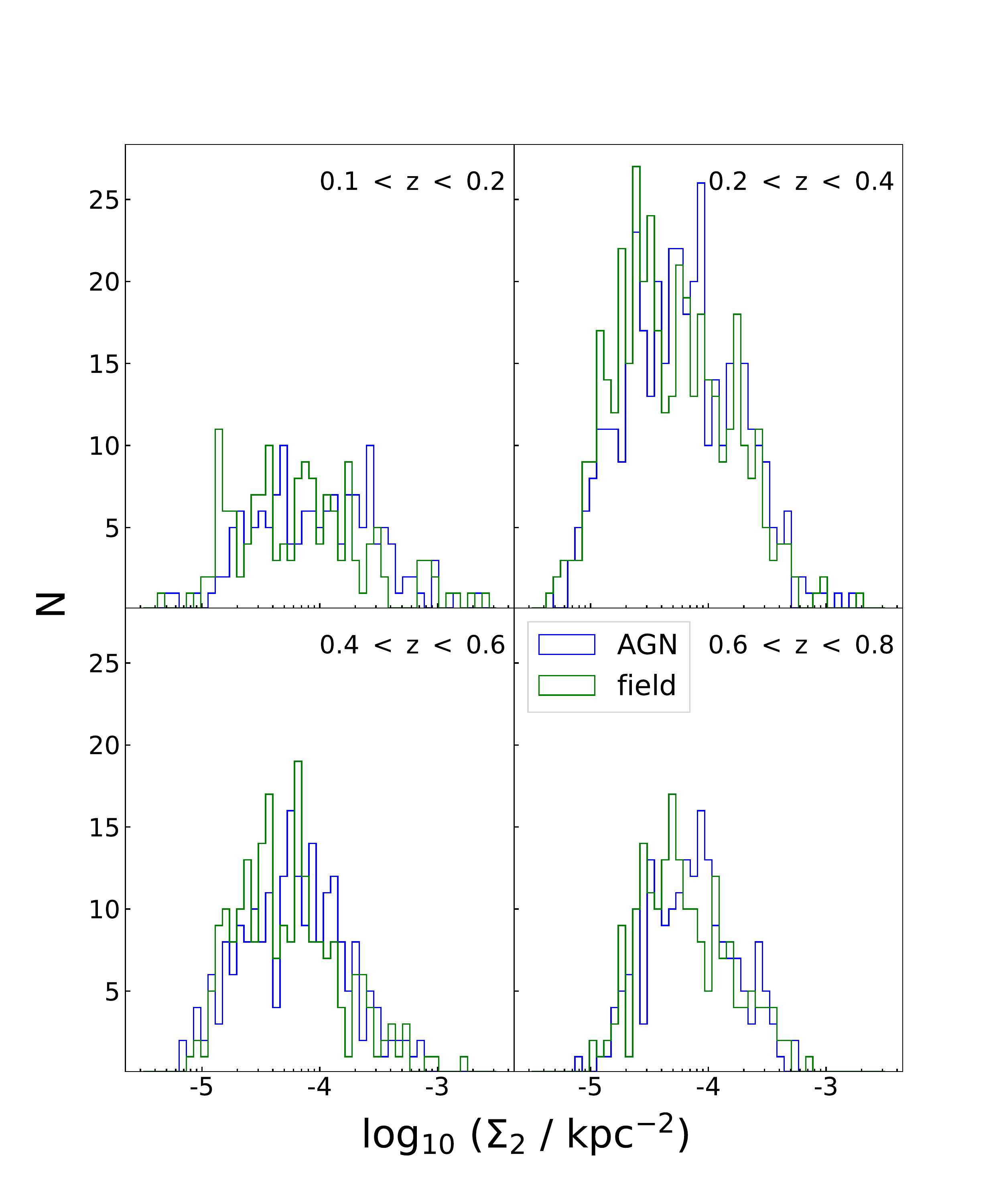}}
 \subfloat[$\Sigma_5$ distributions for AGN and control samples]{\includegraphics[width=0.95\columnwidth]{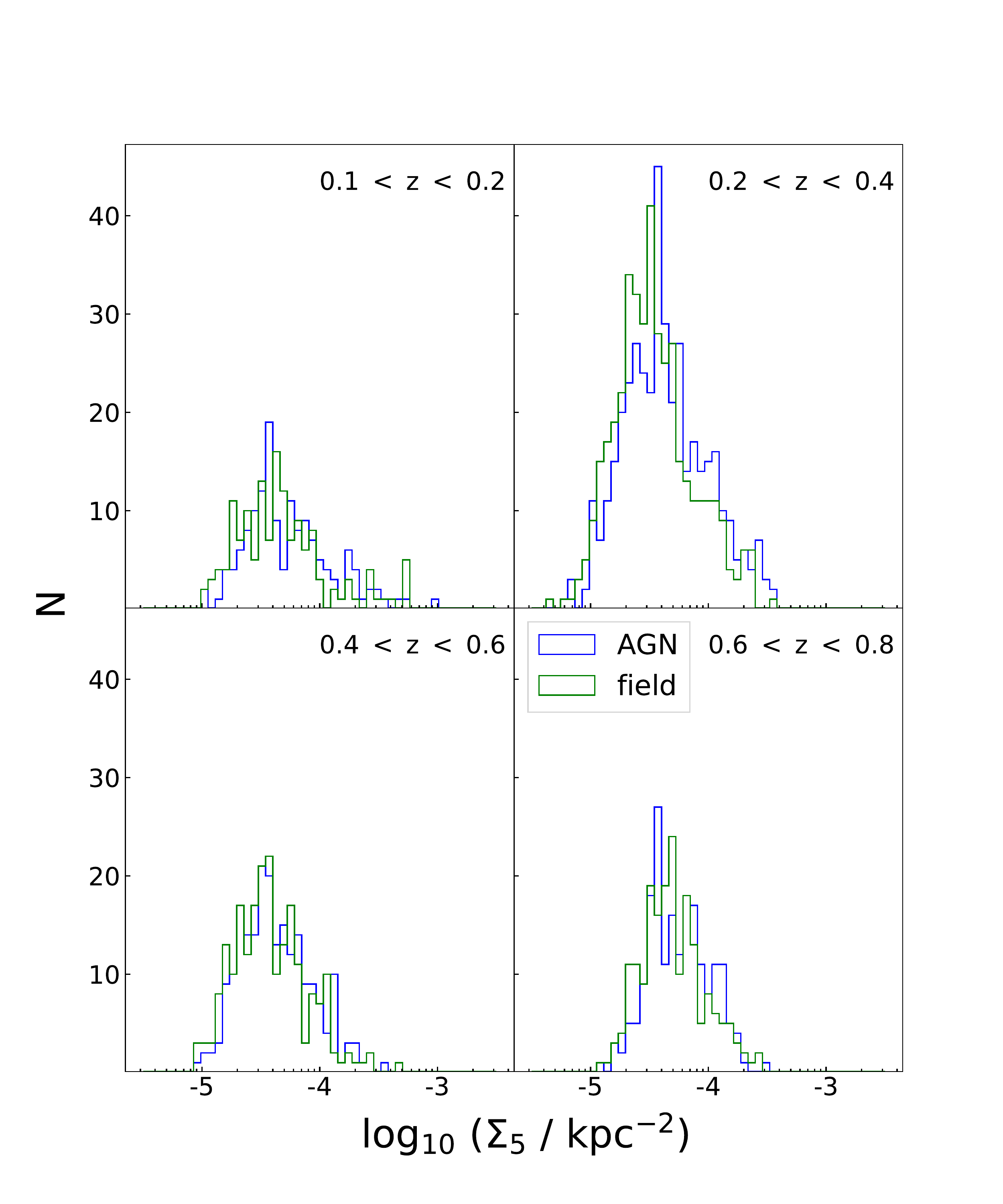}}\\
 \subfloat[$\Sigma_{2,\rm{R}}$ distributions for AGN and control samples]{\includegraphics[width=0.95\columnwidth]{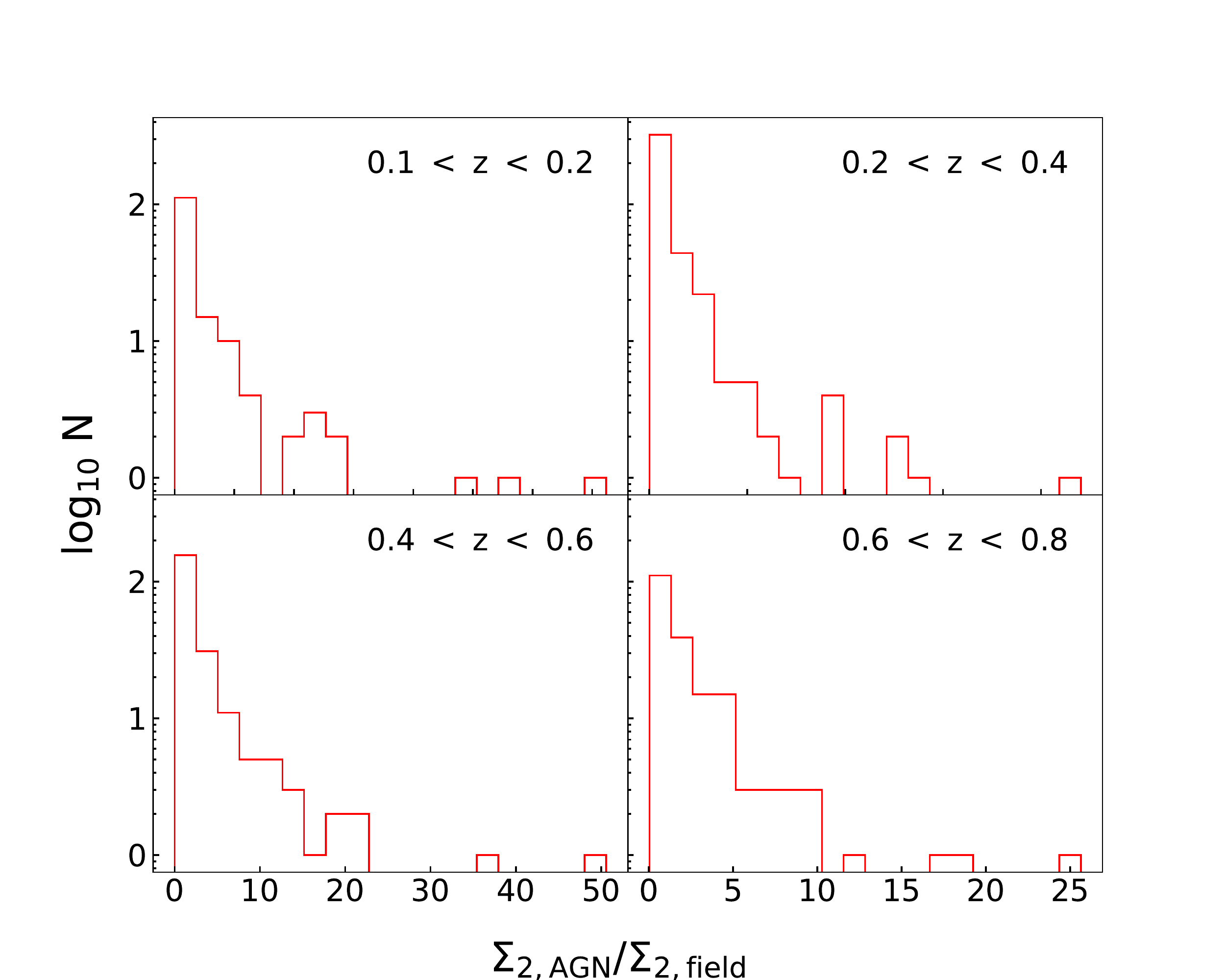}}
 \subfloat[$\Sigma_{5,\rm{R}}$ distributions for AGN and control samples]{\includegraphics[width=0.95\columnwidth]{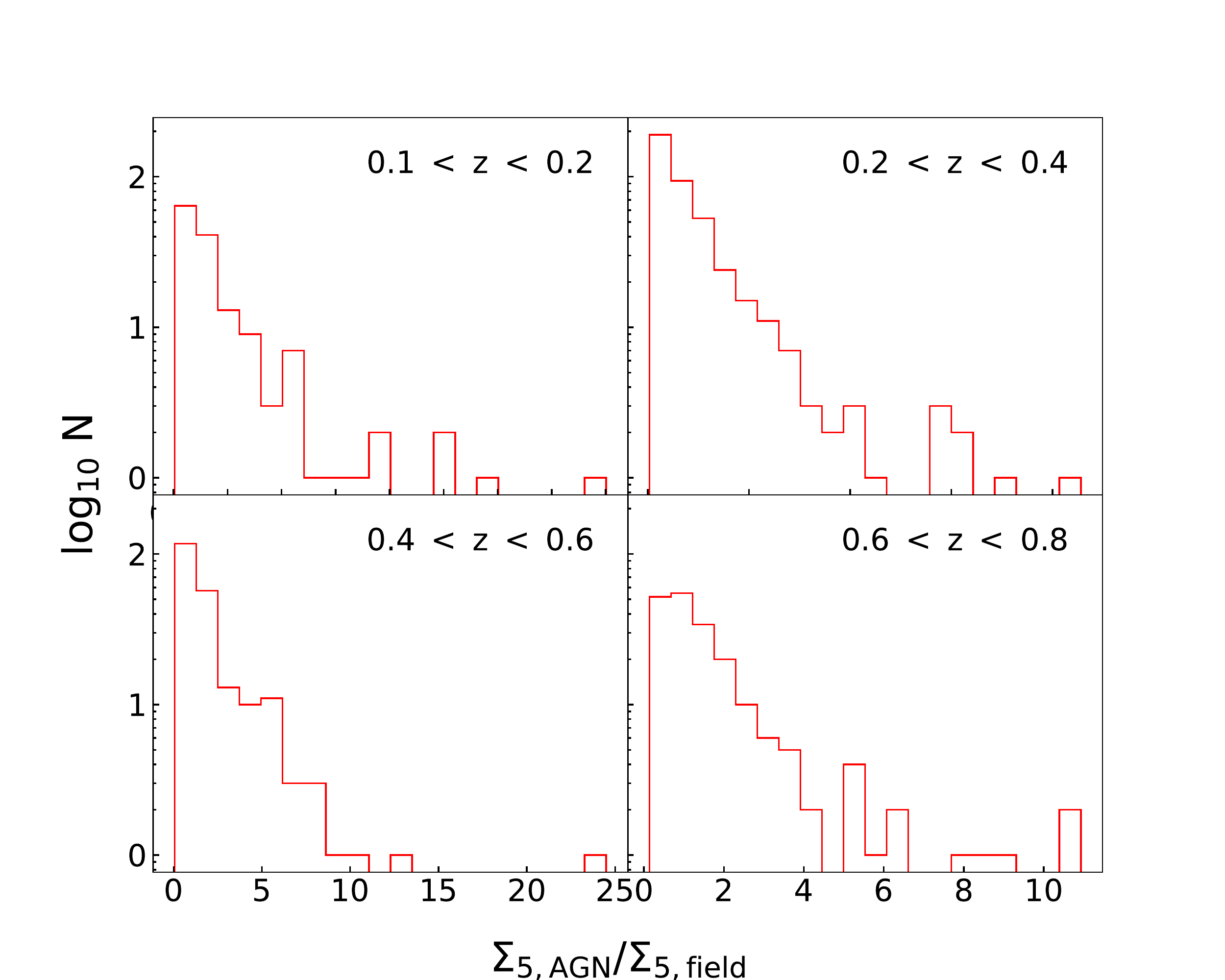}}
  \caption{Distributions of $\Sigma_2$ (kpc$^{-2}$) and $\Sigma_5$ (kpc$^{-2}$) for the AGN and redshift, $K-$band magnitude and ($g-K$)-colour matched control galaxies in panels (a) and (b). The relative density distributions i.e. $\Sigma_{2,\rm{R}}$ and $\Sigma_{5,\rm{R}},$ are also shown in panels (c) and (d).}
  \label{fig:supercontrol-env_L_radio_hist}
\end{figure*}

Galaxy stellar mass and redshift, are not the only issues that may need to be considered in terms of forming a control sample. Galaxy colour may also be related to the galaxy environment \citep[e.g.][]{Madgwick2003,Zehavi2011,Wang2018}. We, therefore, attempt to control for this with sample (iii), which is a redshift, $K-$band magnitude and ($g-K$)-matched sample. Figs.~\ref{fig:K-z-supercontrol} and \ref{fig:g-K-z-supercontrol} show the distributions for the radio source and control sample. 

Fig. \ref{fig:supercontrol-env_L-radio} shows that the mean relative density is $>1$ in $\Sigma_{2,\s{R}}$ and $\Sigma_{5,\s{R}}$ within all redshift and $L_{1.4}$ intervals for control sample (iii). The KS-test results in Table \ref{table:sample3b} indicate that the AGN and control density distributions (both measured using $\Sigma_{2,\s{R}}$ and $\Sigma_{5,\s{R}}$) are inconsistent with being drawn from the same underlying distribution at the $\sim 99$ per cent level.

We also note that the the $p$-values in the KS-tests become progressively larger as we move from sample (i) to sample (iii). As expected the density distributions are most statistically different in the sample that has been matched to the radio AGN with the least number of criteria i.e. redshift only. As the matching procedure becomes more stringent, the statistical difference between the density distributions of AGN and control galaxies becomes lower (see Tables \ref{table:sample1b}, \ref{table:sample2b} and \ref{table:sample3b}).

In Fig.~\ref{fig:supercontrol-env_L_radio_hist}, we also show the distributions of $\Sigma_{2,\s{R}}$ and $\Sigma_{5,\s{R}}$ for both the AGN and control sample (iii). This provides more information on the scales that the environments are being traced. For example, $10^{-5} \lesssim \Sigma_{5,\s{R}} / $\,kpc$^{-2} \lesssim 10^{-3}$ corresponds to radii or separations ($d_N$) of $0.04 \lesssim d_N / \,$kpc $\lesssim 400$. From this we can infer that we are well within the 1-halo term regime, and are only tracing the environments of the AGN and control samples within the dark matter halo in which they reside, and not the larger scale 2-halo clustering.

\begin{table}
\centering
\caption{KS-test results for local AGN density compared to control galaxies (field) density in sample (iii): redshift, $K-$band and ($g-K$)-colour matched field.}
\label{table:sample3b}
\begin{tabular}{c | c c c}
\hline
  & redshifts & $D$ & $p$ \\
 & & & \\
  \hline
    $\frac{\Sigma_{2,{\rm AGN}}}{\Sigma_{2,{\rm field}}}$ 				& 0.1 $<$ z $<$ 0.2  &	0.129  &  9.2\e{-2} \\	
   						 	& 0.2 $<$ z $<$ 0.4  &	0.113  &  4.1\e{-3} \\	
   							& 0.4 $<$ z $<$ 0.6  & 	0.100  &  8.7\e{-2} \\	
   							& 0.6 $<$ z $<$ 0.8  &  0.129  &  1.0\e{-2} \\		
  \hline
  $\frac{\Sigma_{5,{\rm AGN}}}{\Sigma_{5,{\rm field}}}$ 				& 0.1 $<$ z $<$ 0.2  &  0.106  &  2.5\e{-1} \\		
   							& 0.2 $<$ z $<$ 0.4  &  0.136  &  2.6\e{-4} \\		
   							& 0.4 $<$ z $<$ 0.6  &  0.114  &  3.5\e{-2}\\ 		
   							& 0.6 $<$ z $<$ 0.8  &  0.139  &  4.2\e{-3} \\	
  \hline
  \end{tabular}
\end{table}
% -------------
%  section 4
% -------------
\section{Results and Discussion}\label{section-4}

\subsection{Environment and radio luminosity}\label{section:env-jetpower}

Our aim has been to investigate whether there is a link between the galaxy density surrounding AGN with powerful radio emission and the radio luminosity. First, we measure environmental densities of radio AGN relative to the control sample galaxies and investigate whether the environmental density is related to the radio luminosity of the AGN.

From Spearman's rank tests, we find no evidence that the radio luminosity correlates with environment density in both $\Sigma_{2,\s{R}}$ and $\Sigma_{5,\s{R}}$ within the range, $0.1 < z < 0.8$ (see Table \ref{table:correlation_parameters}). This suggests that once triggered the radio luminosity output from the AGN is not directly related to the density of the immediate environment. Thus, at least over the range in radio luminosity that we are able to probe, these radio sources are probably devoid of the hot spots commonly seen in Fanaroff Riley II (FRII)-type radio sources \citep[e.g.][]{FanaroffRiley1974}. Thus radio jets are not halted by the dense intergalactic or intra-cluster medium, which can result in bright hot spots, the luminosity of which tends to scale with the density of the environment \citep[e.g.][]{HardcastleKrause2013}. Instead, the radio sample under investigation here is probably dominated by the much lower luminosity FRI-type source, where there is no sign of a bow shock due to the interaction with the intra-cluster medium, although we still might expect some scaling with environmental density \citep[e.g.][]{LaingBridle2014}

The absence of any correlation between the mean density and $L_{1.4}$ is further proof that radio AGN activity is influenced more strongly by factors other than the galaxy density of radio AGN environments. These can be intrinsic or secular processes such as the accretion rate, temperature and ionisation state of the gas in the interstellar medium surrounding the central nucleus or within the extended gas halo, variations in the polarisation of magnetic fields etc \citep[][e.g]{hardcastle2007a,sabater2015,osullivan2015}.
Given that in order to funnel gas into the central engine, the gas is required to cool, suggests that the amount of cold gas that reaches the central engine is not strongly related to the environmental density that is traced by galaxies.

\subsection{Are the environments of radio AGN different to the field sample?}

Radio-selected AGN commonly trace rich cluster environments \citep[e.g.][]{best2007,karouzos2014a}. Indeed, \citet{magliocchetti2018} find that radio AGN up to $z$ $\sim 1.2$ and with L$_{1.4 \rm{GHz} } > 10^{23.5}$~W~Hz$^{-1}$~sr$^{-1},$ are twice as likely to exist in clusters than their lower luminosity counterparts. The majority (90 per cent) of the sources in our radio AGN sample have radio luminosities above this threshold.  

By performing comparisons with control samples defined in three different ways we are able to investigate how the galaxy mass and colour influence our results. Using just the redshift-matched control sample, we find evidence that the radio sources lie in significant overdensities, however, much of this can be attributed to the fact that they reside in massive host galaxies \citep{Jarvis2001,McLure2004,Seymour2007}, which are the most strongly clustered galaxies. Indeed, when we investigate the environmental density relative to a control sample that is matched in terms of the $K-$band magnitude (a proxy for stellar mass at these redshifts) and redshift, we find that the relative overdensity is significantly reduced but we still find strong evidence that the radio sources reside in overdensities around a factor of two higher than the field.

When we match control galaxies to radio AGN in terms of redshift, $K-$band magnitude and ($g-K$)-colour in sample (iii), we find mixed evidence for a higher environmental density in the radio-AGN sample. In terms of the mean value in each radio luminosity bin, we find significant ($> 3\sigma$) evidence for overdensities in all but the highest radio luminosity bins, for all redshifts. However, if we consider the median overdensity in each bin then the evidence becomes less compelling, with very few bins containing statistically significant overdensities.

This suggests that there is some characteristic of the environmental density that is related to the radio emission, but there is clearly not a strong cause and effect relationship. We suggest that very high density regions are more likely to have a radio AGN present, but that this is certainly not a critical aspect for generating powerful radio emission.

The results of using sample (iii) therefore reveal that the preferred environments of radio AGN are similar to galaxies matched in $K-$band magnitude and colour, but with evidence for a larger proportion of more overdense environments. This finding is supported by results which show that radio-loud AGN are more likely to exist in rich clusters as revealed by observations at z $>$ 1.2 \citep{hatch2011,wylezalek2013} and models and/or simulations \citep{orsi2016,izquierdo-villalba2018}.

Moreover, when controlling for M$_*$ and ($g-K$)-colour, coupled with the fact that we find no correlation between environmental density and radio luminosity. This suggests that the large-scale ($<$1 Mpc) environment is not the only driving factor in triggering and/or maintaining mechanical radio AGN activity. A finding that is in agreement with the work of \cite{sabater2013}, who also find that the prevalence of radio galaxies is only weakly dependent on the large-scale environment.

We, therefore, suggest that although radio AGN, on average, reside in kpc scale overdensities, the triggering and fuelling of the AGN is more dependent on the ability of cold gas to reach the central nucleus. If the total amount of gas is related to the large-scale galactic environment then one might assume that the fuelling of the AGN would also be greater, however we do not find evidence for this. Indeed, the lack of a correlation with radio luminosity and the overdensities around the radio AGN at all radio luminosities and redshifts suggests that the cold gas supply is probably related to something other than the overall environmental density. Possible mechanisms for this are the level and distribution of cold gas in the host galaxy and/or the time since the last merger or interaction which could leave such cold gas in the vicinity of the central black-hole. 

%\newpage
% -------------
% section 5
% ------------
%\twocolumn
\section{Conclusions}\label{section-5}
We have used a 1--2 GHz VLA survey and SDSS Galaxy photometric redshift catalogue to build a radio-selected AGN sample. Using this, we have investigated the role of galaxy density on the radio luminosities of the AGN. By building three control samples: redshift (sample (i)), redshift and $K-$band magnitude matched and redshift (sample (ii)), $K-$band magnitude and ($g-K$)-colour matched (sample (iii)), we investigate the galaxy overdensity for the radio AGN relative to galaxy densities of the matched control sources. We find that close environment densities (at the small group or cluster scale) and radio luminosity are uncorrelated. Thus, this particular scale of galaxy density has a minimal influence on radio AGN power. 

We find that the radio AGN environments are higher in density than the non-AGN matched for all control samples. However, we find that the measured overdensity decreases as we move from using sample (i) to sample (iii) for all redshifts considered. This suggests that both the stellar mass, traced by the $K-$band magnitude, and the host galaxy colour are correlated with the environment, and need to be accounted for when measuring the relative environmental density of AGN.  The magnitude of the overdensities around the radio sources, when compared to the control sample (iii), suggests that radio AGN hosts exist in similar environments to matched samples of non-radio AGN, but that radio AGN are more prevalent in the most overdense regions, such as galaxy clusters.  This suggests that the $<1$ Mpc galaxy environment plays some, but not critical,  role in determining whether a galaxy becomes a radio AGN. Since there is no strong correlation between environment density and $L_{1.4}$, we find no evidence that the environment is strongly linked to the jet power of a radio AGN. To explain this, we suggest that secular processes, such as the rate at which cold gas flows to the nucleus or black-hole mass and spin, have a stronger effect on radio AGN jet emission.

\section*{Acknowledgements}
The authors would like to acknowledge the Department of Science and Technology (DST) and Square Kilometre Array (SKA) of South Africa. This study would not have been possible without funding from National Research Foundation (NRF). We recognise the National Astrophysics and Space Science Program (NASSP) for their teaching. We also thank the University of Western Cape and Oxford University astrophysics departments. We acknowledge Matt Prescott and Imogen Whittam for providing their feedback and opinions on early drafts of this work. The National Radio Astronomy Observatory is thanked for providing data from the Jansky Very Large Array (VLA) as well as SDSS and UKIDSS for their data which is publicly available via \url{http://classic.sdss.org/dr7/} and \url{http://www.ukidss.org/}. 

%---------------------------------------
%			References
%---------------------------------------
\bibliographystyle{mnras}
\bibliography{environment_AGN_stripe82}

\bsp
\label{lastpage}
\end{document}